\documentclass[twocolumn,modern,astrosymb,tighten,trackchanges,article]{aastex631}
\usepackage{comment}
\usepackage{CJK}
\usepackage{amsmath}
\usepackage{appendix}
\usepackage{hyperref}
\usepackage{array,multirow}

\newcommand{\nth}{N$_{2}$H$^{+}$ }

\accepted{17 April 2025}
\submitjournal{The Astrophysical Journal}

\shorttitle{Kinematics of Perseus Protostars}
\shortauthors{Behrens et al.}

\graphicspath{{./}{Figures/}}
\begin{document}
%\begin{CJK*}{UTF8}{gbsn}

\title{Probing the Kinematics of Multiple- and Single-Protostar Systems in Perseus with N$_{2}$H$^{+}$}

   \author[0000-0002-2333-5474]{Erica Behrens}
   \affiliation{Department of Astronomy, University of Virginia, P. O. Box 400325, 530 McCormick Road, Charlottesville, VA 22904-4325}
   \affiliation{National Radio Astronomy Observatory, 520 Edgemont Road, Charlottesville, VA 22903-2475, USA}

   \author[0000-0002-9912-5705]{Adele Plunkett}
   \affiliation{National Radio Astronomy Observatory, 520 Edgemont Road, Charlottesville, VA 22903-2475, USA}

   %\author[0000-0002-9209-7916]{Che-Yu Chen}
   %\affiliation{Lawrence Livermore National Laboratory, Livermore, CA 94550, USA}

   \author[0000-0002-7402-6487]{Zhi-Yun Li}
   \affiliation{Department of Astronomy, University of Virginia, P. O. Box 400325, 530 McCormick Road, Charlottesville, VA 22904-4325}

% Use \input to insert author list here
% Modify standard ALCHEMI author listing to suit specific % article needs
\correspondingauthor{Erica Behrens} \email{eb7he@virginia.edu}

% Use \input to insert email list here
%\input{emaillist.tex}  

%%%%%%%%%%%%%%%%%%%%%%%%%%%%%%%%%%%%%%%%%%%%%%%%%%%%%%%%

\begin{abstract}
We analyze the dense gas kinematics in two Class 0/I protostellar cores, Per\,30 and NGC\,1333 IRAS\,7, in the Perseus molecular cloud to determine whether their velocity structures are indicative of rotation. We examine the hyperfine structure of the \nth $J=1-0$ transition by combining $3^{\prime\prime}$ (900\,AU) Atacama Large Millimeter/Submillimeter Array (ALMA) measurements with $9^{\prime\prime}$ (2700\,AU) measurements from the Green Bank Telescope (GBT). We use the \texttt{CASA Feather} method to combine these data in order to maximize our sensitivity across spatial scales. We fit the \nth spectra to constrain the centroid velocity of the gas at each pixel and use these values to calculate the linear velocity gradient and specific angular momentum within apertures centered on each protostar with radii ranging from $5-60^{\prime\prime}$. Our results indicate that the velocity structure probed by the \nth emission is likely not a result of core rotation. These findings are consistent with other studies in the literature which indicate rotation is often not evident on scales $\lesssim1000$\,AU. We instead suggest that the velocity structure we see is a result of torques caused by irregular density distributions in these protostellar systems.
\end{abstract}

%%%%%%%%%%%%%%%%%%%%%%%%%%%%%%%%%%%%%%%%%%%%%%%%%%%%%%%%

\section{Introduction} \label{sec:intro}

The protostellar phase represents one of the earliest stages of star formation, and the properties of protostellar systems have important implications for the evolution and lives of the resultant stars. One such characteristic is the kinematic structure of protostellar systems {\citep{Offner2023}. Studying protostellar kinematics is a powerful tool for investigating theories of protostar formation as well as understanding the dynamics of multi-star systems. Recent studies reveal that most stars are not formed in isolation \citep{Lada2003,Offner2014}, but it is not yet clear what determines the multiplicity of stellar systems, how the proximity of other protostars affects the evolution of such a system, or how the resulting kinematics would manifest in observations of the molecular gas from which stars form \citep{Krumholz2014,Hacar2023}. The classical picture of star formation suggests that as a dense core forms within a larger-scale molecular cloud, it will collapse, flatten, and begin to rotate \citep{Shu1987}. This rotation and angular momentum may then be inherited by the disk that forms after,
%it should inherit the cloud's angular momentum \textcolor{red}{[CITATION NEEDED]}. Similarly, when disks form from a collapsing core, they too should inherit some portion of the core's angular momentum, 
though this process can be complicated by magnetic braking \citep[e.g.,][]{Li2014}. However, multiple-star systems are likely to feature more complex velocity structure, and it is unclear how each protostar will kinematically influence the dense molecular gas.

%Previous work suggests that as molecular clouds collapse and form stars, the cores and protostellar disks that form inherit the rotation and angular momentum of the larger cloud \textcolor{red}{[NEED REFERENCE--please comment if you know of one]}. 

One such multiple-star system is NGC\,1333~IRAS\,7 (hereafter IRAS\,7) in the nearby Perseus molecular cloud \citep[$d\sim300$\,pc,][]{Tobin2013,Ortiz-Leon2018}, which consists of three protostars: Per\,18, Per\,21, and Per\,49 based on the VLA/ALMA Nascent Disk and Multiplicity Survey \citep{Tobin2016}. More recent observations show that the IRAS\,7 core is in fact a quintuple system, where two of the three main protostellar complexes, Per\,18 and Per\,49, are actually binary systems with separations of 25 \citep{Tobin2016,Reynolds2024} and 94\,AU \citep{Tobin2018}, respectively. Per\,18 and Per\,21 are believed to be Class\,0 sources while Per\,49 is classified as Class\,I \citep{Enoch2009}. All three protostellar systems have known outflows \citep{Tobin2016,Stephens2017}, making this system very kinematically complex.

Previous studies have resolved the disks in all 3 of these protostellar systems \citep{Tobin2016,Tobin2018,Tychoniec2020,Reynolds2024}. Per\,18's two disks are thought to have formed via disk fragmentation as a result of gravitational instability \citep{Tobin2016}. Additionally, \cite{Tobin2018} detect $^{13}$CO, C$^{18}$O, SO, and H$_{2}$CO toward Per\,18, and find that the molecular gas velocity gradients are oriented perpendicular to Per\,18's outflow, indicating that on subarcsecond scales, the kinematics of the Per\,18 system are likely dominated by rotation. Note that \cite{Tobin2016} did not detect any molecular emission toward Per\,49.

%\cite{Chen2019} investigated the velocity structure in several protostellar cores, including IRAS\,7 as well as an isolated protostar Per\,30. \textcolor{red}{[will provide some details here about Per30.]} 
%located in the nearby \citep[$d\sim300$\,pc,][]{Ortiz-Leon2018} Perseus molecular cloud. 
To investigate the kinetic features in star-forming dense cores, the Dynamics in Star-forming cores (DiSCo) survey
\citep{DiSCo2024} used Green Bank Telescope (GBT) measurements of \nth $J=1-0$ at $\sim9''$ (0.01\,pc) spatial resolution to study the dense gas velocity gradients, as probed by \nth $J=1-0$ hyperfine structure. This work suggests that the velocity structure of star-forming cores primarily results from cloud-scale turbulence. In an earlier, more detailed study, \cite{Chen2019} found that on these spatial scales the velocity gradients of the three main protostars in IRAS\,7 (Per\,18, Per\,21, and Per\,49) were aligned nearly parallel to known outflows, which is at odds with the findings of \cite{Tobin2018} using higher-resolution, smaller-scale VLA/ALMA data. If the velocity gradients are interpreted as rotation, this scenario is inconsistent with the classical theory of star formation, where the orientation of angular momentum is conserved from larger to smaller scales \citep{Shu1987}. %and dense cores flatten into rotating disks as a result. 
\cite{Chen2019} suggest the velocity gradient behavior could indicate that rotation, and thus angular momentum, does not always transfer from core to disk scales in a straightforward manner. 

On the other hand, another core in the same molecular complex, Per\,30, showed the opposite picture, with a clear velocity gradient perpendicular to its monopolar outflow \citep{Stephens2018}. Per\,30 is an isolated single protostellar system with smooth kinetic features and a candidate disk identified by \cite{Segura-Cox2018}.
These results indicate that obtaining measurements with resolution elements between the $9^{\prime\prime}$ beam from \cite{Chen2019} and the subarcsecond beam of \cite{Tobin2018} is necessary in order to understand the origin of velocity gradients and thoroughly disentangle the complex velocity structure seen in multiple-star versus isolated systems.

In order to bridge the gap in resolution seen in previous studies, it is imperative that we consider data that is sensitive to spatial scales in between those presented in \cite{Tobin2018} and \cite{Chen2019}. We must consider kinematic behavior across scales in order to understand the inheritance of velocity structure and angular momentum from protostellar cores to disks. In this paper we present intermediate-resolution measurements from the Atacama Large Millimeter/submillimeter Array (ALMA) of the \nth $J=1-0$ transition and its hyperfine lines for the IRAS\,7 and Per\,30 cores. We combine these data with the coarser-resolution GBT data from \cite{Chen2019} in order to increase the range of scales to which we are sensitive with our measurements. 

The ALMA data, their combination with GBT data, and our efforts to spectrally fit the data in order to extract velocity information is described in Section \ref{sec:data}. In Section \ref{sec:results} we present integrated intensity and centroid velocity maps of the \nth $J=1-0$ emission. An analysis and discussion of these results and their implications for the evolution of protostellar systems can be found in Section \ref{sec:discussion}.

\section{DATA} \label{sec:data}

\begin{deluxetable*}{ccccccccccccc} 
\tablecaption{Measurement details for the ALMA, GBT, and Feather data. \label{tab:rms}}
\setlength{\tabcolsep}{3pt}
    \tablehead{ & \multicolumn{4}{c}{ALMA} & & \multicolumn{3}{c}{GBT} & & \multicolumn{3}{c}{Feather} \\
    \cline{2-5} \cline{7-9} \cline{11-13}
     \colhead{Field} & \colhead{continuum,} & \colhead{$v_{\text{res}}$} & \colhead{$\theta_{\text{maj}} \times \theta_{\text{min}}$} & \colhead{LAS} & & 
     \colhead{\nth} & \colhead{$v_{\text{res}}$} &  \colhead{$\theta_{\text{maj}} \times \theta_{\text{min}}$} & & \colhead{\nth} & \colhead{$v_{\text{res}}$} &  \colhead{$\theta_{\text{maj}} \times \theta_{\text{min}}$} \\ [-5pt]
     & \nth RMS & & & &    & RMS& & &    &RMS & & \\
     & (mJy/beam) & (km/s) & ($''$) & ($''$) & &
     (Jy/beam) & (km/s) & ($''$) & &
     (mJy/beam) & (km/s) & ($''$)  
     }
    \startdata
    Per\,18-21 & 0.5, 17.5 & 0.05 & $3.93 \times 2.73$ & 33.1 &  & 0.60 & 0.023 & 9.$40 \times 9.40$ & & 18.2 & 0.05 & $3.93 \times 2.73$\\
    Per\,49 & 0.4, 18.0 & 0.05 & $3.95 \times 2.72$ & 33.2 &  & 0.60 & 0.023 & 9.$40 \times 9.40$ & & 18.0 & 0.05 & $3.95 \times 2.72$ \\
    Per\,30 & 0.4, 17.3 & 0.05 & $3.93 \times 2.72$ & 33.0 & & 0.65 & 0.023 & 9.$40 \times 9.40$ & & 19.8 & 0.05 & $3.93 \times 2.72$ \\
    \enddata
\end{deluxetable*}

\subsection{Observations} \label{sec:obs}

\begin{figure}
    \gridline{\fig{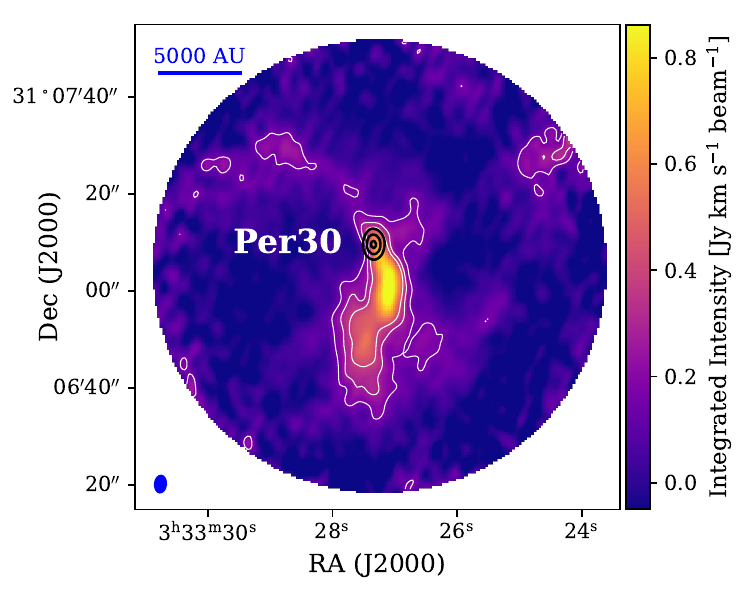}{0.47\textwidth}{(a)}}
    \vspace{-3mm}
    \gridline{\fig{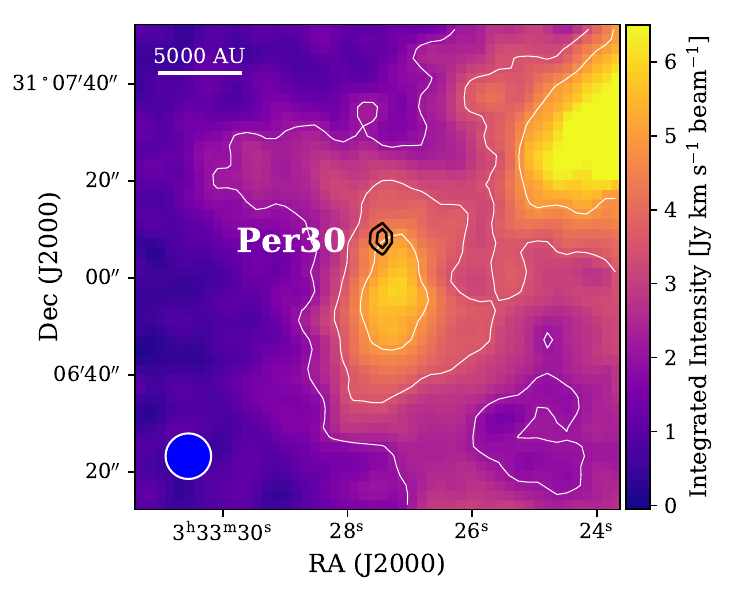}{0.47\textwidth}{(b)}}
    \vspace{-3mm}
    \gridline{\fig{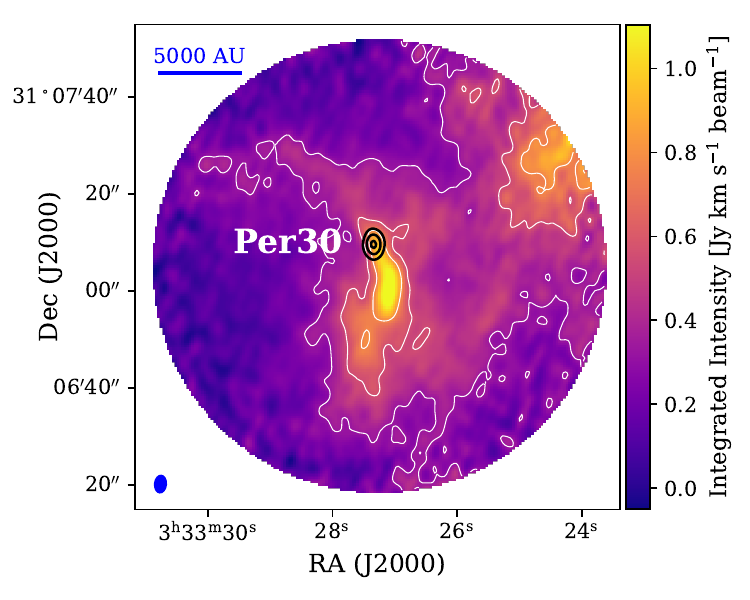}{0.47\textwidth}{(c)}}
    \vspace{-3mm}
    \caption{Integrated intensity maps of the Per\,30 field from (a) ALMA-only data, (b) GBT-only data, and (c) Feather data. Black contours represent ALMA 93\,GHz continuum emission at the 5, 10, and 15\,$\sigma$ levels (see Table \ref{tab:rms} for $\sigma$ values). White contours indicate the integrated intensity 3, 5, and 7\,$\sigma$ levels. The beam sizes for each image are represented by blue ellipses in the bottom left corner. }
    \label{fig:per30_mom0}
\end{figure}

%\begin{figure*}
%    \gridline{\leftfig{Per30_alma_mom0_noCut.pdf}{0.5\textwidth}{(a)}
%    \rightfig{Per30_gbt_mom0_noCut.pdf}{0.5\textwidth}{(b)}}
%    \vspace{-5mm}
%    \gridline{\fig{Per30_feather_mom0_noCut.pdf}{0.5\textwidth}{(c)}}
%    \vspace{-2mm}
%    \caption{Integrated intensity maps of the Per\,30 field from (a) ALMA-only data, (b) GBT-only data, and (c) Feather data. Black contours represent ALMA 93\,GHz continuum emission at the 5, 10, and 15\,$\sigma$ levels. White contours indicate the integrated intensity 3, 5, and 7\,$\sigma$ levels. See Table \ref{tab:rms} for exact $\sigma$ values. The beam sizes for each image are represented by blue ellipses in the bottom lefthand corner. }
%    \label{fig:per30_mom0}
%\end{figure*}

\begin{figure}
    \gridline{\fig{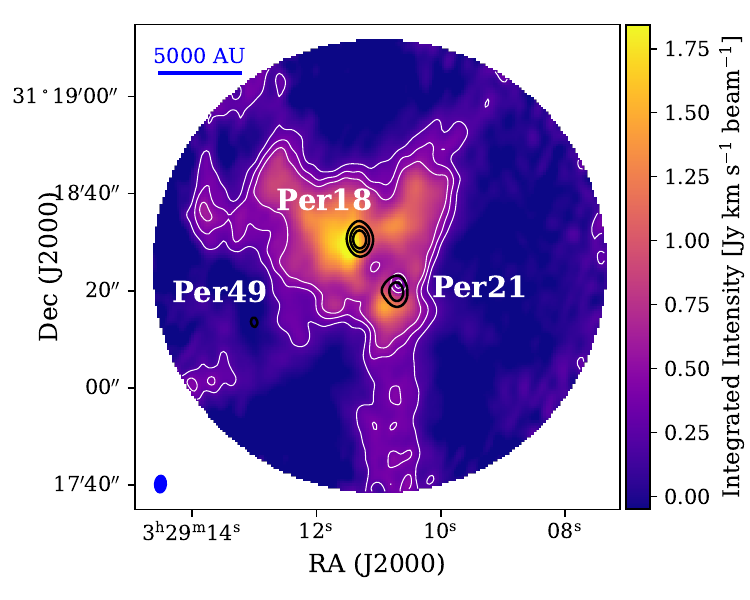}{0.47\textwidth}{(a)}}
    \vspace{-3mm}
    \gridline{\fig{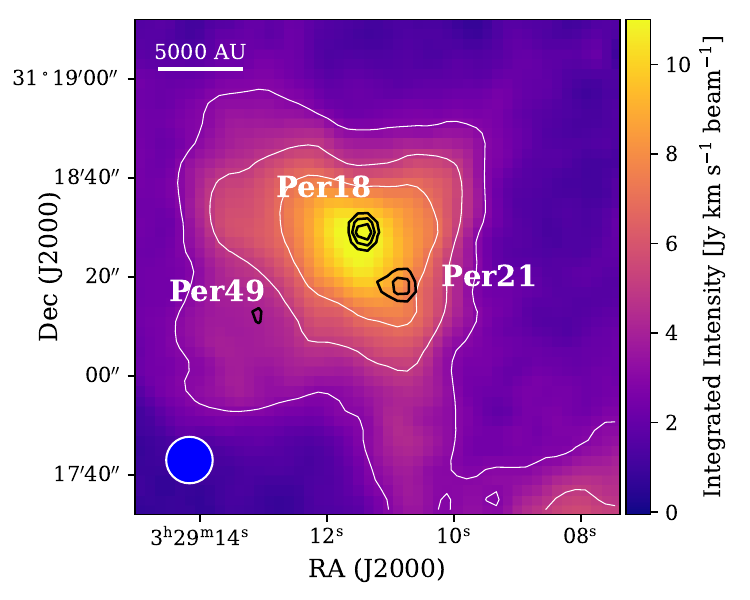}{0.47\textwidth}{(b)}}
    \vspace{-3mm}
    \gridline{\fig{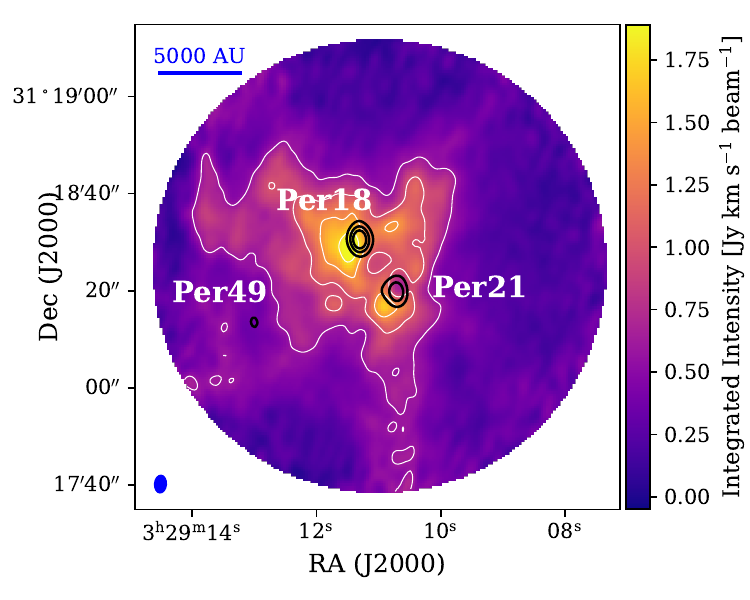}{0.47\textwidth}{(c)}}
    \vspace{-3mm}
    \caption{Same as in Figure \ref{fig:per30_mom0} but for the field centered on Per\,18-21 in the IRAS\,7 core. Per\,18 is represented by the northernmost contours, Per\,21 is located just southwest of Per\,18, and the location of Per\,49 can be seen in the southeast near the edge of the field.}
    \label{fig:per18-21_mom0}
\end{figure}

%\begin{figure*}
%    \gridline{\leftfig{Per18-21_alma_mom0_noCut.pdf}{0.5\textwidth}{(a)}
%    \rightfig{Per18-21_gbt_mom0_noCut.pdf}{0.5\textwidth}{(b)}}
%    \vspace{-5mm}
%    \gridline{\fig{Per18-21_feather_mom0_noCut.pdf}{0.5\textwidth}{(c)}}
%    \vspace{-2mm}
%    \caption{Same as in Figure \ref{fig:per30_mom0} but for the field centered on Per\,18-21 in the IRAS\,7 core. Per\,18 is represented by the northernmost contours, Per\,21 is located just southwest of Per\,18, and Per\,49 can be seen in the southeast near the edge of the field.}
%    \label{fig:per18-21_mom0}
%\end{figure*}

\begin{figure}
    \gridline{\fig{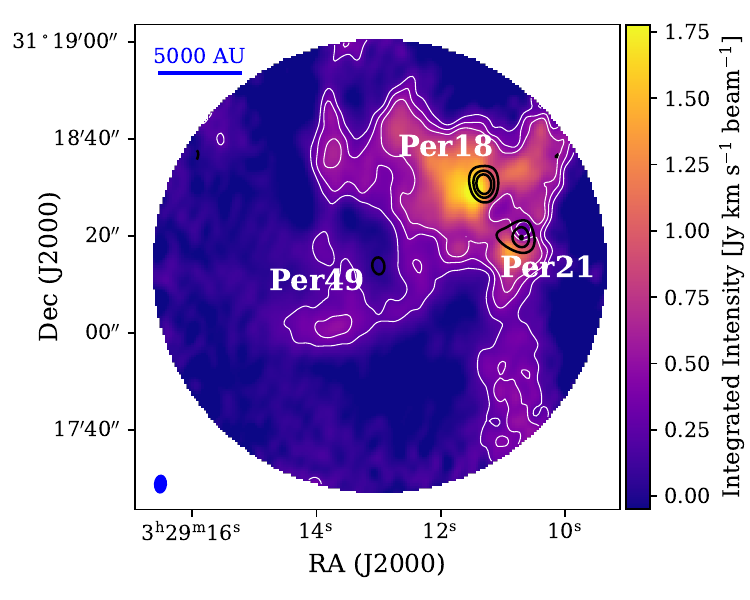}{0.47\textwidth}{(a)}}
    \vspace{-3mm}
    \gridline{\fig{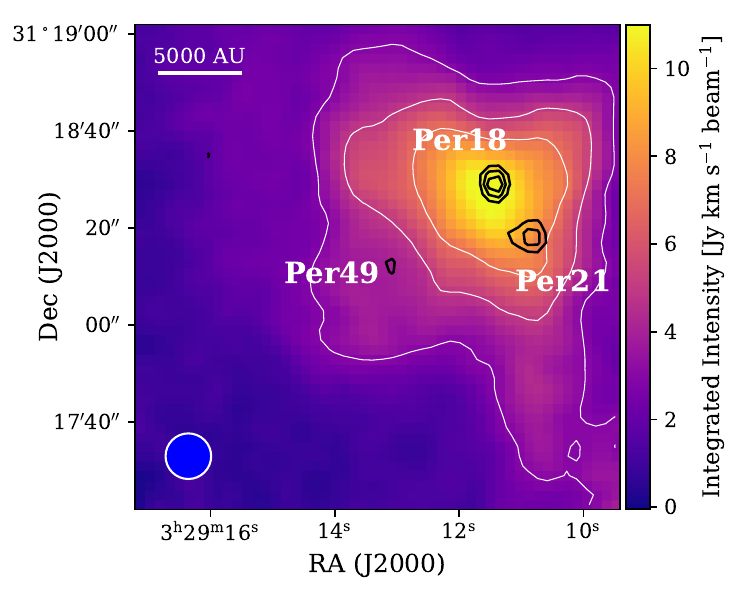}{0.47\textwidth}{(b)}}
    \vspace{-3mm}
    \gridline{\fig{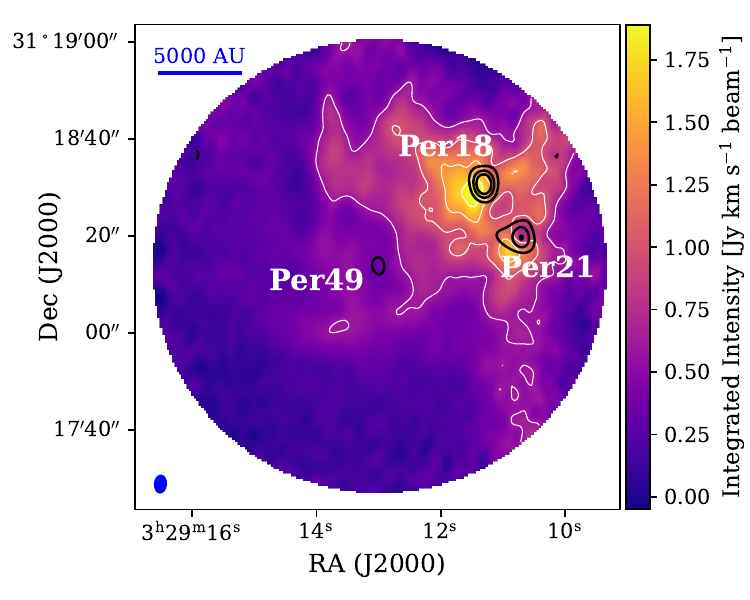}{0.47\textwidth}{(c)}}
    \vspace{-3mm}
    \caption{Same as in Figure \ref{fig:per18-21_mom0} but for the IRAS\,7 field centered on Per\,49.}
    \label{fig:per49_mom0}
\end{figure}

%\begin{figure*}
%    \gridline{\leftfig{Per49_alma_mom0_noCut.pdf}{0.5\textwidth}{(a)}
%    \rightfig{Per49_gbt_mom0_noCut.pdf}{0.5\textwidth}{(b)}}
%    \vspace{-5mm}
%    \gridline{\fig{Per49_feather_mom0_noCut.pdf}{0.5\textwidth}{(c)}}
%    \caption{Same as in Figure \ref{fig:per18-21_mom0} but for the IRAS\,7 field centered on Per\,49.}
%    \label{fig:per49_mom0}
%\end{figure*}

We target the N$_{2}$H$^{+}$ $J=1-0$ transition using data from ALMA project 2019.1.00914.S (P.I. C. Chen) toward two cores in the Perseus molecular cloud, Per\,30 and IRAS\,7. These observations were taken December 24-25, 2019 on the 12m array using 47 antennas and tuned to a rest frequency of 93.173402 GHz (sky frequency 93.170745 GHz). These data were calibrated using the automated ALMA calibration pipeline, where source J0423-0120 was used for bandpass and flux calibration, and J0336+3218 was used for phase calibration. We reduced these data using the radio astronomy data reduction package \texttt{CASA} \citep{CASATeam2022} following standard ALMA reduction procedures, where we used natural weighting when cleaning the continuum data and briggs weighting with a robust value of 0.5 for spectral line imaging. We divide our data into three fields, with one centered on Per\,30 (Figure \ref{fig:per30_mom0}, phase center $\alpha$(ICRS) = $03^h33^m27^s.220$, $\delta$(ICRS) = +$31^{\circ}07^{\prime}05^{\prime\prime}.680$) and the other two (overlapping) fields centered on the IRAS\,7 protostars Per\,18-21 (Figure \ref{fig:per18-21_mom0}, phase center $\alpha$(ICRS) = $03^h29^m10^s.960$, $\delta$(ICRS) = +$31^{\circ}18^{\prime}25^{\prime\prime}.630$) and Per\,49 (Figure \ref{fig:per49_mom0}, phase center $\alpha$(ICRS) = $03^h29^m12^s.957$, $\delta$(ICRS) = +$31^{\circ}18^{\prime}14^{\prime\prime}.307$), respectively. The final synthesized beamsizes are 3.$''$9 $\times$ 2.$''$7 with velocity resolution of 0.05\,km\,s$^{-1}$ and a largest angular scale of $\sim33''$.

We combine these data with lower-resolution \nth $J=1-0$ data of the same cores from the Green Bank Telescope (GBT, see \cite{Chen2019} for more details on these observations). These  data have a beam size of 9.$''$4 and 1.43 kHz (0.023 km\,s$^{-1}$) spectral resolution. Details of both the ALMA and GBT observations can be found in Table \ref{tab:rms}.

%\begin{figure*}
%    \centering
%    \includegraphics[trim = 10mm 0mm 5mm 0mm, scale=0.8]{GBT+feather+ALMA_gbtRes.pdf}
%    \caption{N$_{2}$H$^{+}$ $J=1-0$ spectra from the GBT (blue), ALMA (black), and combined using Feather(red), all smoothed to GBT resolution of 9.$''$4.}
%    \label{fig:compare_feather}
%\end{figure*}

\begin{figure}
    \gridline{\fig{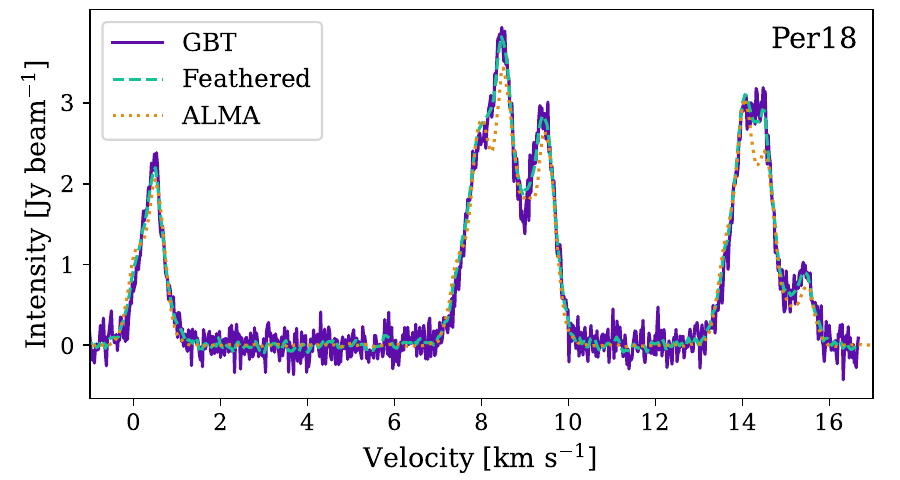}{0.47\textwidth}{(a)}}
    \vspace{-3mm}
    \gridline{\fig{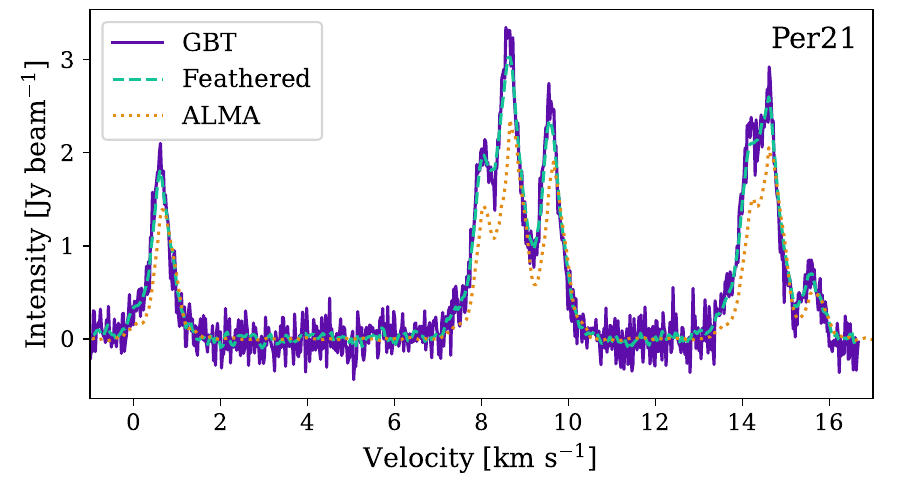}{0.47\textwidth}{(b)}}
    \vspace{-3mm}
    \gridline{\fig{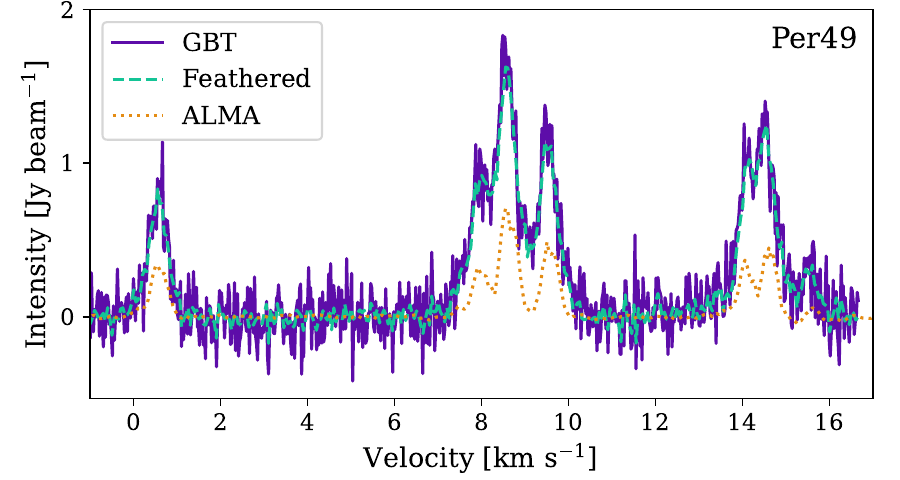}{0.47\textwidth}{(c)}}
    \vspace{-3mm}
    \gridline{\fig{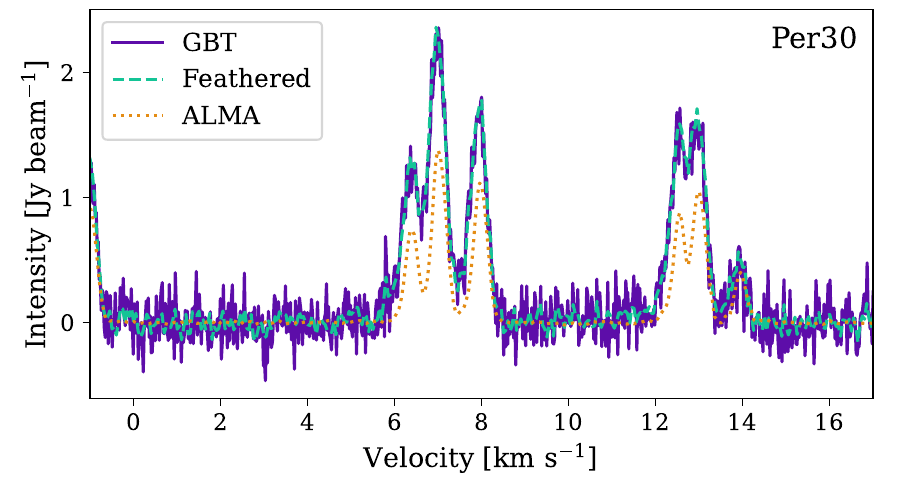}{0.47\textwidth}{(d)}}
    \vspace{-3mm}
    \caption{Spectra from the locations of the Per\,18 (a), Per\,21 (b), Per\,49 (c) and Per\,30 protostellar systems from ALMA, GBT, and Feather data, smoothed to a common resolution of 9.$''$4.}
    \label{fig:spectra}
\end{figure}

%\begin{figure*}
%    \centering
%    \includegraphics[scale=0.65, trim=15mm 3mm 17mm 5mm, clip=True]{Per18-21_gbt_feather_alma_spectrum.pdf}
%    \caption{Spectra from the location of the Per\,18 protostar from ALMA, GBT, and Feather data, smoothed to a common resolution of 9.$''$4.}
%    \label{fig:Per18spec}
%\end{figure*}

%\begin{figure*}
%    \centering
%    \includegraphics[scale=0.65, trim=8mm 3mm 17mm 5mm, clip=True]{Per21_gbt_feather_alma_spectrum.pdf}
%    \caption{Spectra from the location of the Per\,21 protostar from ALMA, GBT, and Feather data, smoothed to a common resolution of 9.$''$4.}
%    \label{fig:Per21spec}
%\end{figure*}

%\begin{figure*}
%    \centering
%    \includegraphics[scale=0.65, trim=9mm 3mm 17mm 5mm, clip=True]{Per49_gbt_feather_alma_spectrum.pdf}
%    \caption{Spectra from the location of the Per\,49 protostar from ALMA, GBT, and Feather data, smoothed to a common resolution of 9.$''$4.}
%    \label{fig:Per49spec}\end{figure*}

\begin{figure}
    \gridline{\fig{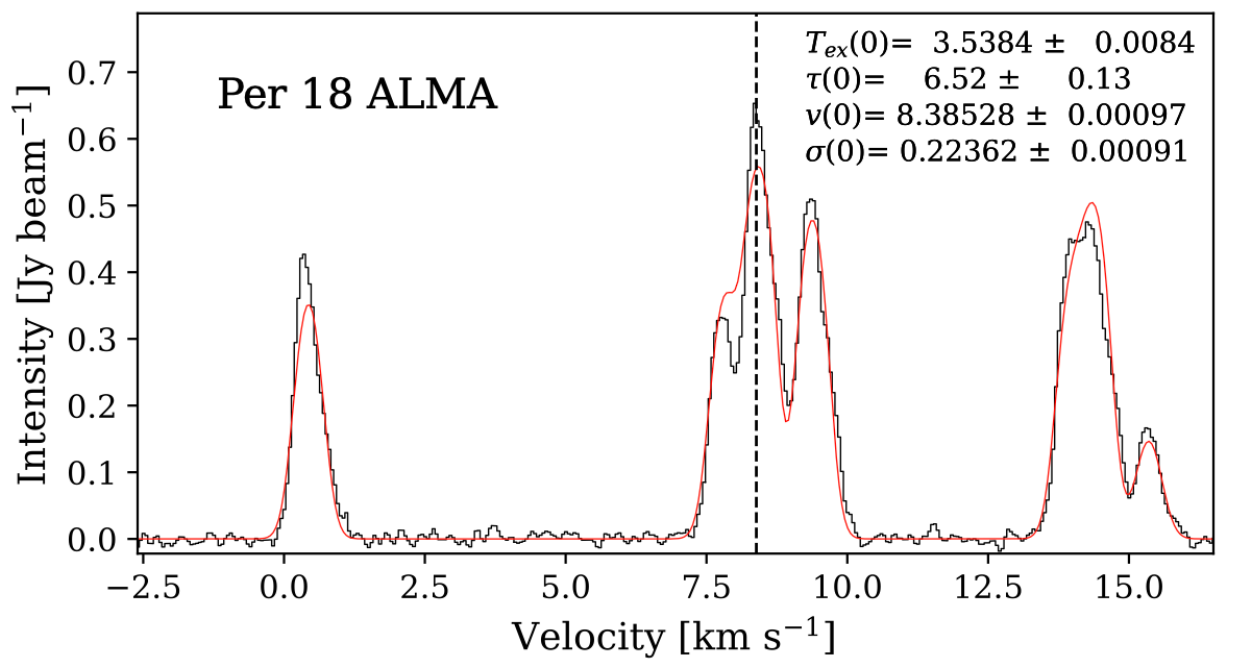}{0.47\textwidth}{(a)}}
    \vspace{-3mm}
    \gridline{\fig{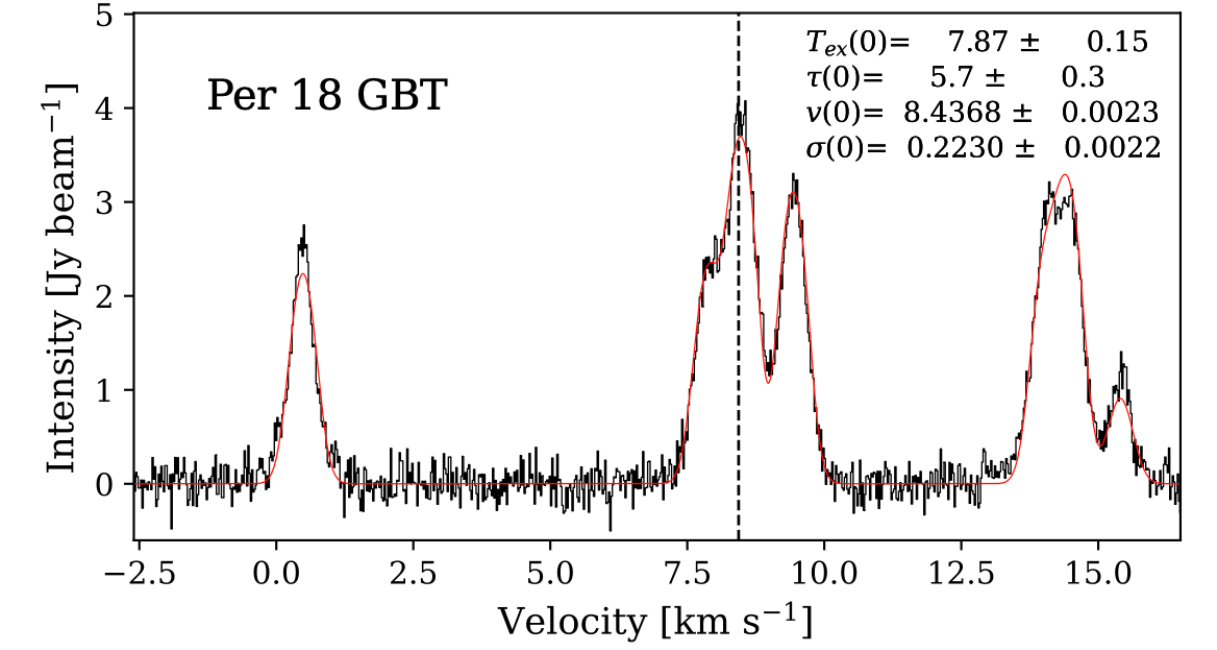}{0.47\textwidth}{(b)}}
    \vspace{-3mm}
    \gridline{\fig{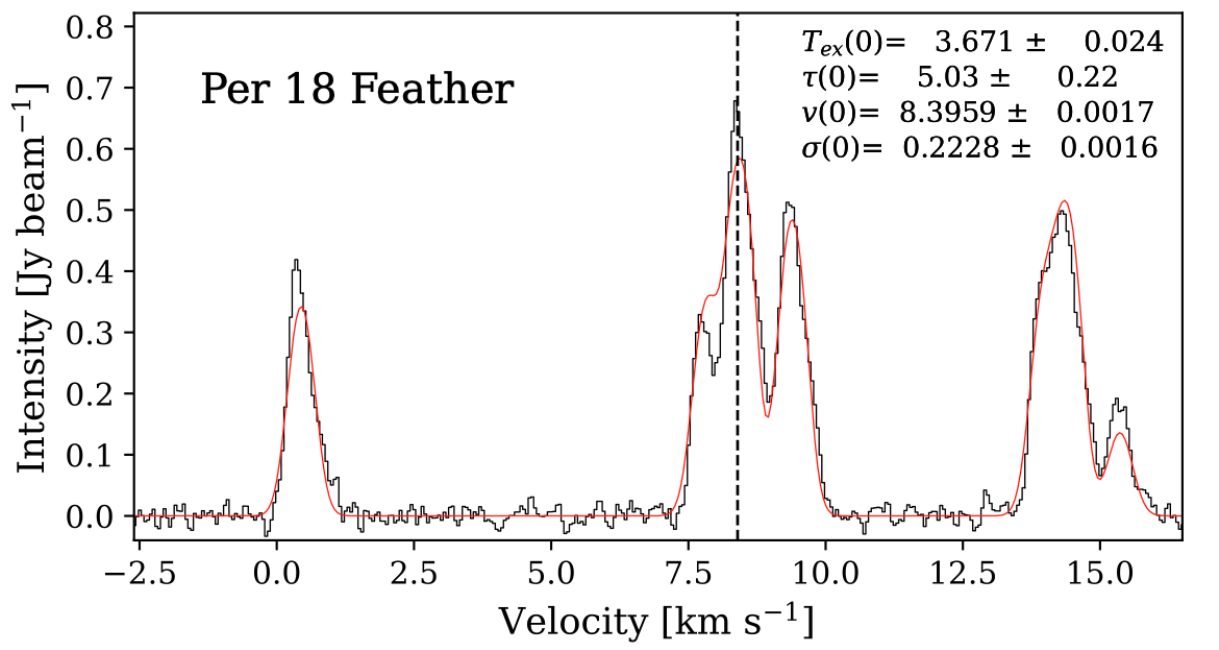}{0.47\textwidth}{(c)}}
    \vspace{-3mm}
    \caption{Sample \nth $J=1-0$ spectral fits (red curves) of the ALMA (a), GBT (b), and Feather (c) spectra (solid black curves) toward the ALMA Per\,18 continuum peak. Each panel shows the best-fit values and uncertainties for excitation temperature $T_\text{ex}$ in K, the total optical depth $\tau$ (which is distributed across the 15 hyperfine components), centroid velocity $v$ (also denoted by the dashed black lines) in km\,s$^{-1}$, and linewidth $\sigma$ in km\,s$^{-1}$.}
    \label{fig:specFits}
\end{figure}

%\begin{figure*}
%    \centering
%    \includegraphics[scale=0.65, trim=10mm 0 18mm 12mm, clip=True]{Per18_feather_specFit.png}
%    \caption{A sample of the \nth $J=1-0$ Feather spectrum (black) near the Per\,18 protostar, overlaid by the fitted model (red) from \texttt{pyspeckit}. Fitted parameter values and uncertainties are displayed in the legend.}
%    \label{fig:Per18-21_specFit}
%\end{figure*}

%\begin{figure*}
%    \centering
%    \includegraphics[scale=0.65, trim=10mm 0 15mm 30, clip=True]{Per21_feather_specFit.png}
%    \caption{Same as in Figure \ref{fig:Per18-21_specFit} but for the gas around protostar Per\,21.}
%    \label{fig:Per21_specFit}
%\end{figure*}

\subsection{Data Combination} \label{sec:data_comb}

We use the \texttt{CASA} method \texttt{feather}\footnote{\url{https://casa.nrao.edu/docs/taskref/feather-task.html}} to combine our interferometry and single-dish data. \texttt{feather} combines the baselines of these two images in the Fourier plane, allowing us to examine both the large- and small-scale gas structure in these cores. This combination allows us to benefit from both ALMA's higher spatial resolution as well as the GBT's sensitivity to larger spatial scales. Before feathering the data, we regrid the GBT data to match our ALMA images and then multiply these regridded images by ALMA's primary beam response, following the prescription laid out in \cite{Hoffman2018}. We adopt a single-dish scaling factor of 1.0, which indicates that at spatial frequencies where there is overlap in the GBT and ALMA images, the emission in the two datasets is equal. We then provide the \texttt{feather} method with the ALMA and regridded GBT images to create a feathered image with information from all of the available baselines.

In order to determine whether our Feather attempt was sufficient, we compare the Feather, ALMA, and GBT spectra smoothed to a common resolution of 9.$''$4 at each protostar location. Since the ALMA data is limited by the spatial scales dictated by its antenna baselines, it will only contain a fraction of the emission measured by the GBT ($\lesssim50$\% for Per\,30 and Per\,49 and $\gtrsim75$\% for Per\,18-21), which contains the total emission from the region. Thus, an ideal Feather spectrum will be equal or nearly equal in intensity to the single-dish spectrum; a Feather spectrum with lower intensity than that of the single-dish would indicate that not all single-dish flux has been recovered. As we see in Figure \ref{fig:spectra} for Per\,18, our Feather spectra align well with our single-dish spectra, indicating we have recovered most or all of the available flux. Note that though the feathered images should provide us with the truest picture of the gas emission around these protostellar systems, we will continue to consider the ALMA and GBT data as well in order to better understand how characteristics of observations can influence data interpretation.

\subsection{Spectral Fitting} \label{sec:specFit}
To analyze the velocity structure of these protostars, we extract the \nth $J=1-0$ spectra from our ALMA, GBT, and Feather data cubes and fit the spectra in order to estimate the centroid velocity and linewidth of the spectrum at each pixel. We use the Python package \texttt{PySpecKit}\footnote{\url{https://pyspeckit.readthedocs.io/en/latest/index.html}} \citep{Ginsburg2011,Ginsburg2022} and its N$_{2}$H$^{+}$ hyperfine fitter to fit the hyperfine components of the $J=1-0$ transition in each of our datasets. We adopt a signal-to-noise ratio (SNR) cutoff of $5\sigma$, where $\sigma$ is the single-channel RMS of signal-free channels (see Table \ref{tab:rms})---below this cutoff, \texttt{PySpecKit} will not attempt to fit the spectra due to a lack of sufficient signal. We also test using an SNR cutoff of $3\sigma$ to determine how low-signal emission affects the spectral fits, which will be discussed more in Section~\ref{sec:cenV_maps}. For those pixels that do meet our SNR criteria, \texttt{PySpecKit} provides estimates and uncertainties associated with the excitation temperature, the total optical depth from the contributions of the 15 hyperfine components, the centroid velocity, and the linewidth at each pixel. Figure \ref{fig:specFits} shows the results of the spectral fitting routine for ALMA, GBT, and Feather data toward Per\,18, including fitted parameter values and uncertainties, where the centroid velocity uncertainties are on the order of $10^{-3}$~km\,s$^{-1}$. We also provide our spectral fits toward the three remaining protostars in Appendix~\ref{sec:allFits}.

Since other similar studies, albeit of larger systems, have shown two velocity components in dense gas tracers like \nth and NH$_{3}$ \citep[e.g.][]{Hacar2017,Chen2022}, we test our spectral fit with two components. For the case of the Per\,18-21 ALMA data, focusing toward Per\,18, \texttt{PySpecKit} recovers two components, one with a centroid velocity of 8.01\,km\,s$^{-1}$ and another with a centroid velocity of 8.35\,km\,s$^{-1}$. The velocity of this second component is within 1\% of the centroid velocity we retrieve with a one component fit toward Per\,18, and the excitation temperature (3.61\,K) is within 2\% of the single-component excitation temperature we use here (see Figure~\ref{fig:specFits}a). Additionally, \texttt{PySpecKit} finds that the excitation temperature of the 8.01\,km\,s$^{-1}$ component in the two-component fit is $10^{-5}$\,K. Thus we conclude that if a second component does indeed exist, its contribution to the \nth emission is minimal. As a result we assume that the \nth emission in these cores is primarily tracing a single component. Our use of single-component fits is also consistent with previous spectral fitting of the \nth GBT measurements performed in \cite{Chen2019}.

\section{Results} \label{sec:results}

\subsection{Integrated Emission Morphology} \label{sec:mom0_maps}

We examine the total \nth emission by considering moment-0 integrated intensity maps. Figures \ref{fig:per30_mom0}, \ref{fig:per18-21_mom0}, and \ref{fig:per49_mom0} show the moment-0 maps for the Per\,30, Per\,18-21, and Per\,49 fields, respectively, derived from our ALMA, GBT, and Feather datasets. Based on the location of the continuum contours, we see that the peak of the \nth integrated emission is consistently offset from the dust continuum in all three of our fields, but most notably in Per\,30. \cite{Chen2019} comments on this offset, noting that the gas immediately around the protostar may be warmer, resulting in more gaseous CO which will then combine with the \nth gas to form HCO$^{+}$ \citep{Tobin2016,van'tHoff2017}. However, this process likely occurs on smaller scales ($\lesssim0.01$\,pc, or 7$''$) than the separation between the dust continuum and \nth peaks ($\sim10''$). This chemical pathway is also likely why we see a dearth of \nth gas around the Per\,49 protostar in Figure \ref{fig:per49_mom0}, especially in the smaller-scale images from ALMA. While Per\,18 and Per\,21 are class 0 sources, Per\,49 is a more evolved class I source; thus we expect Per\,49 to have less envelope material that can be traced by \nth. 

Comparing the \nth emission between our ALMA, GBT, and Feather datasets, we observe that the ALMA data features more compact emission as a result of the longer interferometric baselines being sensitive to smaller- rather than larger-scale structures. The most significant ALMA \nth emission is clustered toward the centers of the fields, whereas the GBT captures more diffuse emission that extends farther away from the protostars (see, for example the emission in the upper right of Figure~\ref{fig:per30_mom0}b). Since the Feather images should capture the total \nth emission across all spatial scales, we can see contributions from both the high-resolution compact ALMA emission toward the centers of the fields, as well as the more diffuse GBT emission, which extends to the edges of our ALMA fields.

\subsection{Centroid Velocity Maps} \label{sec:cenV_maps}

\begin{figure}
    \gridline{\fig{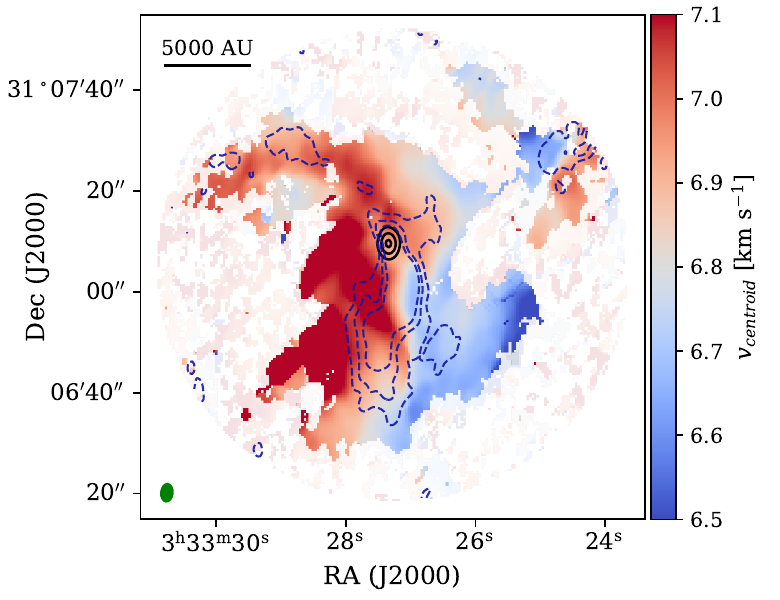}{0.48\textwidth}{(a)}}
    \vspace{-3mm}
    \gridline{\fig{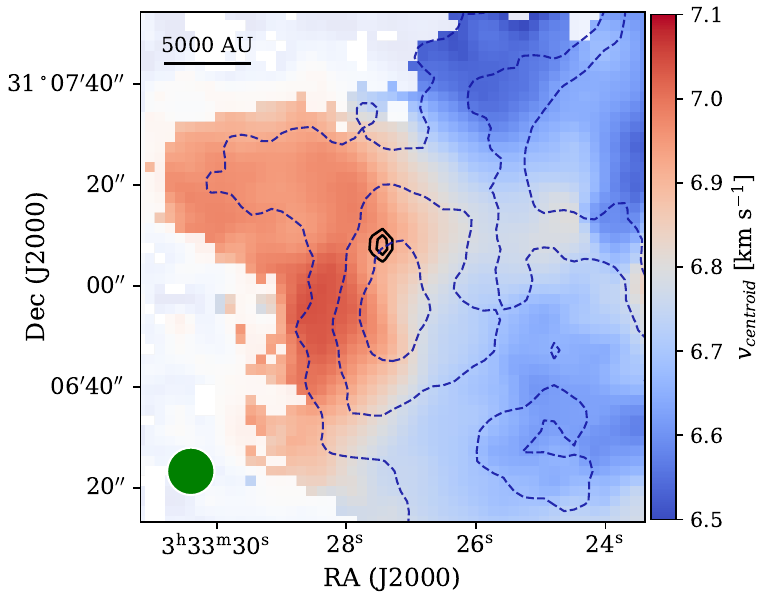}{0.48\textwidth}{(b)}}
    \vspace{-3mm}
    \gridline{\fig{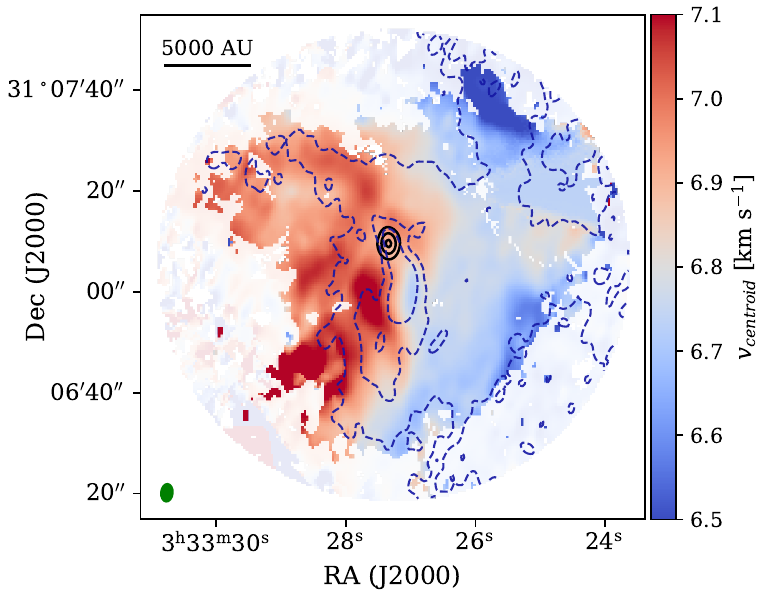}{0.48\textwidth}{(c)}}
    \vspace{-3mm}
    \caption{Fitted centroid velocity maps of Per\,30 derived from spectral fits of our (a) ALMA-only, (b) GBT-only, and (c) Feather datasets. The translucent and opaque coloring denotes areas where the integrated intensity SNR $\geq$ 3 and SNR $\geq$ 5, respectively. The black solid contours are the same as in Figure~\ref{fig:per30_mom0}, showing the dust continuum emission, and the blue dashed contours are the same as the white contours in Figure~\ref{fig:per30_mom0}, showing the integrated \nth emission.}
    \label{fig:per30_cenv}
\end{figure}

%\vspace{-5mm}
%\begin{figure*}
%    \gridline{\leftfig{Per30_centroidv_alma.pdf}{0.5\textwidth}{(a)}
%    \rightfig{Per30_centroidv_gbt.pdf}{0.5\textwidth}{(b)}}
%    \vspace{-5mm}
%    \gridline{\fig{Per30_centroidv_feather.pdf}{0.5\textwidth}{(c)}}
%    \caption{Fitted centroid velocity maps of Per\,30 derived from spectral fits of our (a) ALMA-only, (b) GBT-only, and (c) Feather datasets.}
%    \label{fig:per30_cenv}
%\end{figure*}

%\vspace{-20mm}
\begin{figure}
    \gridline{\fig{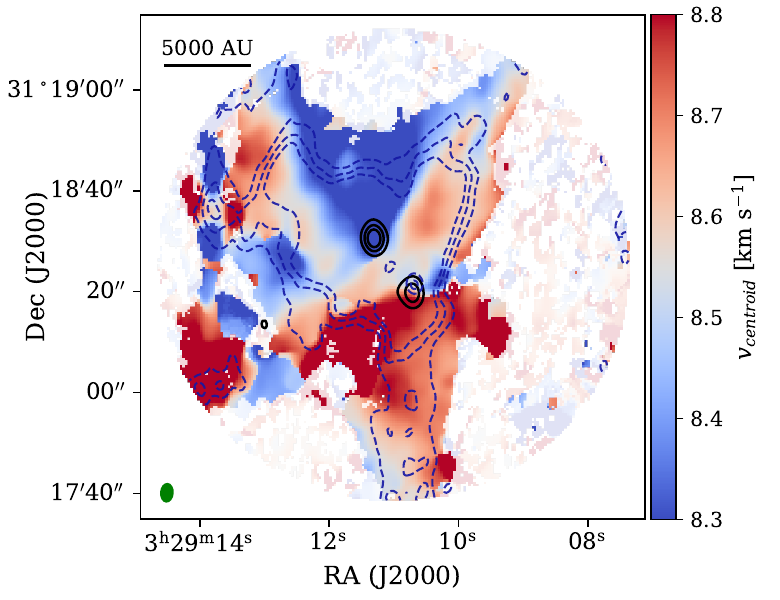}{0.48\textwidth}{(a)}}
    \vspace{-3mm}
    \gridline{\fig{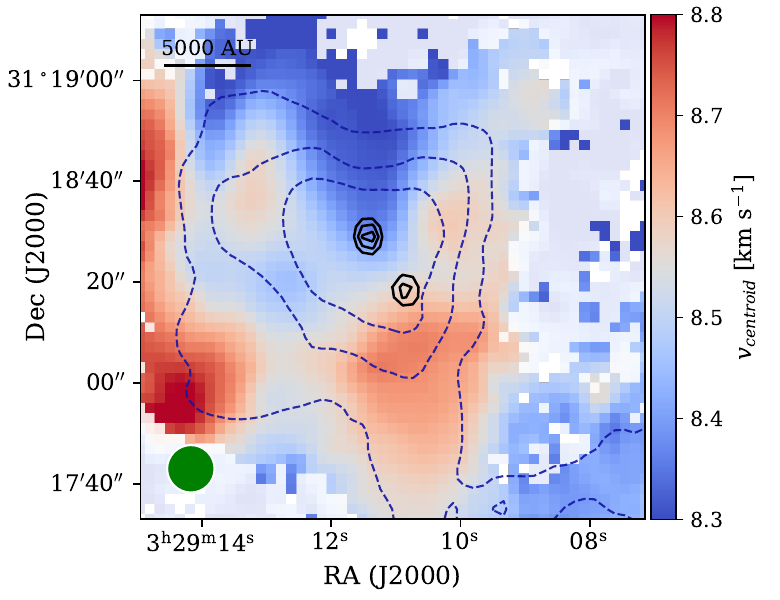}{0.48\textwidth}{(b)}}
    \vspace{-3mm}
    \gridline{\fig{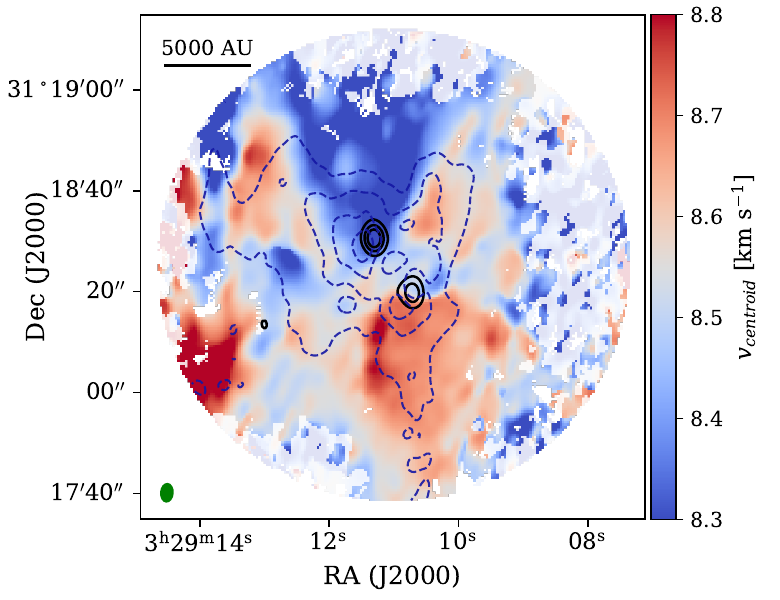}{0.48\textwidth}{(c)}}
    \vspace{-3mm}
    \caption{Fitted centroid velocity maps of Per\,18-21 derived from spectral fits of our (a) ALMA-only, (b) GBT-only, and (c) Feather datasets. The translucent and opaque coloring denotes areas where the integrated intensity SNR $\geq$ 3 and SNR $\geq$ 5, respectively.}   
    \label{fig:per18-21_cenv}
\end{figure}

\begin{figure}
    \gridline{\fig{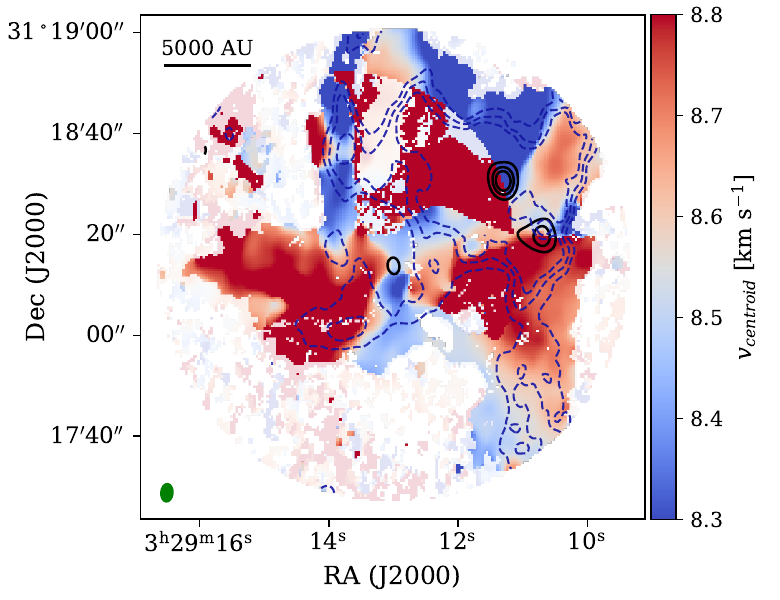}{0.48\textwidth}{(a)}}
    \vspace{-3mm}
    \gridline{\fig{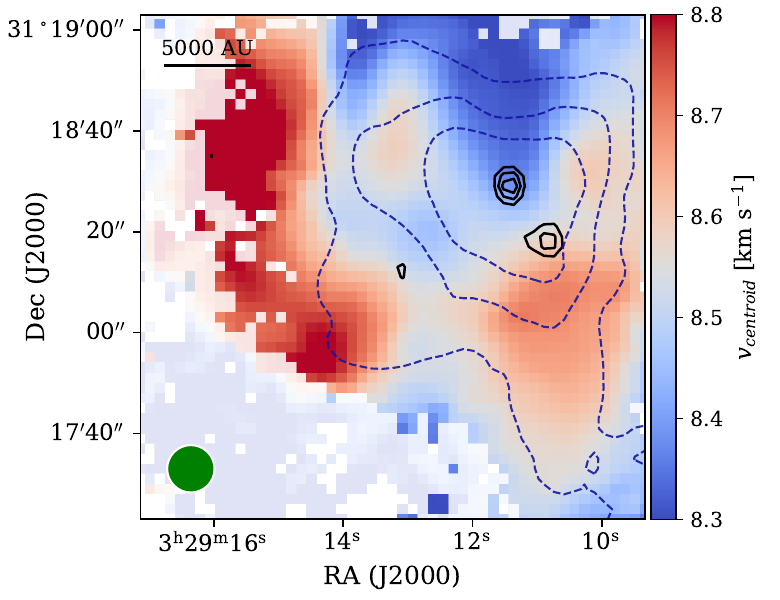}{0.48\textwidth}{(b)}}
    \vspace{-3mm}
    \gridline{\fig{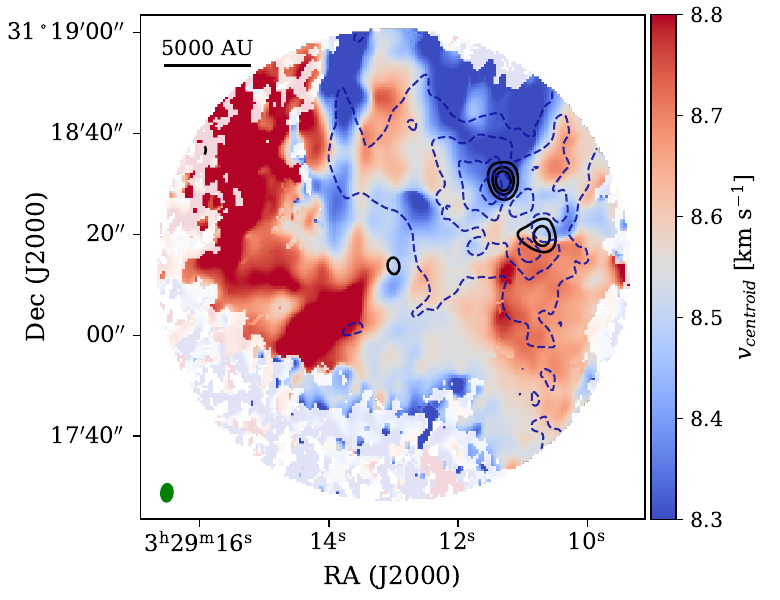}{0.48\textwidth}{(c)}}
    \vspace{-3mm}
    \caption{Fitted centroid velocity maps of Per\,49 (center) and Per\,18-21 (right) derived from spectral fits of our (a) ALMA-only, (b) GBT-only, and (c) Feather datasets. The translucent and opaque coloring denotes areas where the integrated intensity SNR $\geq$ 3 and SNR $\geq$ 5, respectively. }   
    \label{fig:per49_cenv}
\end{figure}

To investigate the velocity structure of the \nth emission in our three fields, we use our fitted centroid velocity values from our spectral fitting routine (see Section \ref{sec:specFit}) to make maps of the centroid velocity across these fields. Figures \ref{fig:per30_cenv}, \ref{fig:per18-21_cenv}, and \ref{fig:per49_cenv} show the centroid velocity maps for our Per\,30, Per\,18-21, and Per\,49 fields, respectively, derived from the ALMA, GBT, and Feather datasets. Emission that meets our $5\sigma$ SNR threshold is shown with opaque coloring, and emission where $3\leq$ SNR $\leq5$ is shown in translucent coloring. Since using only a $3\sigma$ SNR cutoff resulted in the inclusion of many noisy pixels near the edges of the fields, particularly in the case of the ALMA and feather images, we use the results associated with the $5\sigma$ SNR cutoff for the remainder of our analysis.

These centroid velocity maps figures demonstrate distinct velocity structure at different spatial scales. In general, the ALMA centroid velocity maps feature more pronounced velocity gradients, especially nearer to the protostar. 
%We also see patchiness in the background of these images as a result of our fitting routine either ignoring or struggling to fit the spectra in low signal-to-noise regions. Since the background areas contain mostly diffuse gas, which is invisible to ALMA at our chosen baselines, little signal from smaller-scale gas exists there, resulting in poor or patchy spectral fits. 
Conversely, our GBT and Feather centroid velocity maps show smoother velocity structure extending farther away from the protostar. Though the GBT data captures gas across all size scales, the intensity of the diffuse gas in combination with the GBT's resolution makes the GBT maps ideal for analyzing large-scale structure. Our GBT measurements effectively capture the more diffuse gas, revealing the velocity structure of this less dense component. Incorporating the GBT data allows us to  obtain better spectral fits farther from the protostars, which we are unable to achieve using ALMA data alone.

In the Per\,30 field (Figure \ref{fig:per30_cenv}), we see the most likely indications of rotation among our four protostars. The Per\,30 maps demonstrate a clear transition from redshifted to blueshifted gas from east to west across the field of view, which is consistent with other works \citep[e.g.][]{Goodman1993} that have attributed velocity gradients in protostellar cores to rotation. Our ALMA data also reveals a potential streamer-like structure (see the top panels of Figures \ref{fig:per30_mom0} and \ref{fig:per30_cenv}), redshifted and coming from the east and north of the protostar. While the GBT and Feather images also show redshifted \nth gas in the same area of the IRAS\,7 field, the ALMA centroid velocity map shows a clearly-defined stream of velocity-coherent gas arcing in toward the protostar. Additionally, while all three datasets show a transition from redshifted to blueshifted gas southwest of the protostar, the ALMA data indicates the smaller-scale gas in this region is moving slightly faster and is more redshifted than the larger-scale gas probed by the GBT. While modeling the velocity gradient and possible rotation that would be indicative of a streamer as described by \citet{Pineda2023} is beyond the scope of this work, further analysis may provide more clues to the importance of this feature for accretion onto the source.

In the IRAS\,7 fields, the ALMA centroid velocity maps (Figures \ref{fig:per18-21_cenv}a and \ref{fig:per49_cenv}a) again demonstrate the most complex velocity structure. Both figures feature a blueshifted, V-shaped structure to the north of Per\,18. Though this feature is also present in the GBT and Feather centroid velocity maps, the ALMA maps show a more clearly defined transition from redshifted to blueshifted gas on either side of the ``V". This structure is somewhat aligned with Per\,18's outflow, which has a position angle of $345^{\circ}$ \citep{Tobin2016}. Other notable features that are most prominent in the IRAS\,7 ALMA centroid velocity maps are the redshifted wing-like structures extending east and west away from Per\,49, seen most notably in Figure \ref{fig:per49_cenv}. While we also see these redshifted features in the GBT and Feather images, the smaller scales gas is moving $0.1-0.2$\,km\,s$^{-1}$ faster than the more diffuse gas. Per\,49's outflow has a position angle of $\sim27^\circ$ \citep{Stephens2017}, and thus does not appear to be aligned with this feature. On the other hand, the GBT centroid velocity maps capture a large clump of redshifted gas northeast of Per\,49 that is less pronounced in the ALMA maps. It is clear from these maps that the kinematic nature of small- and large-scale gas structures varies greatly on scales of $\gtrsim 1000$\,AU.

\subsection{Linear Velocity Gradients}\label{sec:lvgrad}

In order to address the question of rotation in these protostellar systems, we consider the specific angular momentum $J_{\textrm{core}}$ of these cores across spatial scales. Previous observational \citep{Hsieh2021} and theoretical \citep{Misugi2023} studies have suggested that cores inherit angular momentum from their larger-scale progenitors, namely filaments and clouds. We will examine the behavior of angular momentum across scales in our protostellar systems by considering this quantity on large and small spatial scales with our ALMA and GBT data. Additionally, examining the conservation, or lack thereof, of angular momentum for a given core by calculating $J_{\textrm{core}}$ at various distances from its center allows us to assess whether the velocity structure we measure is truly due to rotation.  

To probe specific angular momentum, we must first calculate the linear velocity gradients $\nabla v_{\textrm{lsr}}$ around each protostar. $\nabla v_{\textrm{lsr}}$ is related to $J_{\textrm{core}}$ by $J_{\textrm{core}} \equiv L_{\textrm{core}} / M_{\textrm{core}} = r_{\textrm{core}}^{2} \nabla v_{\textrm{lsr}}$, where $L_{\textrm{core}}$ and $M_{\textrm{core}}$ are the angular momentum and the mass of the core, respectively \citep{Goodman1993,Chen2019}. We consider $\nabla v_{\textrm{lsr}}$ over circular regions, or apertures, centered around each protostar (using coordinates provided in \cite{Tobin2016}) with radii ranging from 5--30$''$. We choose the lower end of this range to ensure that each region is at least one ALMA beam width across, and we consider distances no greater than 30$''$ in order to avoid confusion between the IRAS\,7 protostars. In each region, we use the centroid velocity maps presented in Section \ref{sec:cenV_maps} to calculate the velocity gradient (both its magnitude and direction) at each pixel using the NumPy \texttt{gradient} function\footnote{\url{https://numpy.org/doc/2.1/reference/generated/numpy.gradient.html}} and then average together the gradient values for all pixels within a given region. Note that this method is the same as that used in \cite{Chen2019} but differs from those used in some other studies which derive linear velocity gradients from gas around protostars \citep[e.g.][]{Gaudel2020,Sai2023}. We derive uncertainties on $\nabla v_{\textrm{lsr}}$ by calculating a standard deviation on $\nabla v_{\textrm{lsr}}$ for all the pixels within a given region. We repeat this procedure for the ALMA, GBT, and Feather datasets, the values for which are listed in Table \ref{tab:lvgrad}. We successfully replicate the $\nabla v_{\textrm{lsr}}$ values calculated for IRAS\,7 GBT data in \cite{Chen2019} for 10$''$ circular regions centered on each of the three IRAS\,7 protostars. Since \cite{Chen2019} use an alternative core definition for Per\,30 based on Herschel continuum contours, our $\nabla v_{\textrm{lsr}}$ values calculated for a 10$''$ region are not fully comparable. We then use our linear velocity gradient values to calculate angular momentum at various radii using the relation above. We plot angular momentum as a function of radii for each protostar and image type in Figure \ref{fig:specAngMom}, which we will discuss further in Section \ref{sec:rotation} below.

\begin{deluxetable*}{cccccccccc}
\centering
\tablecolumns{10}
\tablecaption{Linear velocity gradients calculated over circular regions centered on each protostar with radii ranging from 5--30$''$. \label{tab:lvgrad}}
\tablehead{
 & & \multicolumn{2}{c}{ALMA} & & \multicolumn{2}{c}{GBT} & &  \multicolumn{2}{c}{Feather} \\ \cline{3-4} \cline{6-7} \cline{9-10}
\colhead{Protostar} & \colhead{Radius} & \colhead{$\nabla v_{lsr}$} & \colhead{$\theta_{\nabla v_{lsr}}$} & &\colhead{$\nabla v_{lsr}$} & \colhead{$\theta_{\nabla v_{lsr}}$} & &\colhead{$\nabla v_{lsr}$} & \colhead{$\theta_{\nabla v_{lsr}}$} \\ 
 & ($''$) & (km\,s$^{-1}$\,pc$^{-1}$) & ($^{\circ}$) & & (km\,s$^{-1}$\,pc$^{-1}$) & ($^{\circ}$) & & (km\,s$^{-1}$\,pc$^{-1}$) & ($^{\circ}$) 
}
\startdata
Per\,30 & 5  &  $ 25.1\pm 18.8$ &  85.4  & &  $9.9\pm2.4$ &   83.5 & & $17.5\pm13.9$ &   84.6 \\
        & 10 &   $107.6\pm700.4$ &  101.0  & &  $8.7\pm2.9$ &   82.6 & & $12.1\pm13.1$ &   84.8 \\
        & 15 &   $98.5\pm 817.3$ &  107.1 & &  $7.2\pm3.2$ &   78.4 & &  $7.6\pm11.8$ &   77.6 \\
        & 20 &   $36.5\pm969.8$&  146.6 & &  $5.6\pm3.2$ &   78.4 & &  $3.5\pm317.6$ &   -7.6 \\
        & 25 &   $27.9\pm1426.6$&  66.1 & &  $8.2\pm120.3$ &   -82.4 & &  $25.2\pm697.8$ &   -60.4 \\
        & 30 &   $61.3\pm1628.0$ &   65.3 & &  $33.6\pm268.1$ &  -74.0 & &  $60.4\pm1104.2$ &   3.7 \\ \hline 
Per\,18 &  5 &  $24.0\pm32.3$ & -157.6 & &  $7.6\pm2.4$ &  177.0 & & $14.7\pm13.4$ & -165.3 \\
        & 10 &  $11.3\pm26.1$ & -157.3 & &  $6.3\pm4.2$ & -166.0 & &  $8.8\pm15.4$ & -163.8 \\
        & 15 &   $8.2\pm25.6$ & -164.8 & &  $5.2\pm4.5$ & -155.5 & &  $4.7\pm15.4$ & -163.0 \\
        & 20 &   $30.6\pm618.9$ & 91.86 & &  $4.2\pm4.3$ & -160.5 & &  $5.6\pm17.1$ & -159.1 \\
        & 25 &  $68.4\pm1058.4$ & 125.3 & &  $3.6\pm4.2$ & -165.6 & &  $7.3\pm190.6$ & -177.9 \\
        & 30 &  $61.6\pm1498.0$ &  120.6 & &  $19.2\pm191.9$ & 105.2 & &  $20.9\pm630.9$ &  154.3 \\ \hline
Per\,21 &  5 &  $24.4\pm46.6$ & 161.9 & &  $9.3\pm1.1$ & -161.3 & & $18.5\pm26.0$ &  166.4 \\
        & 10 &  $14.9\pm804.7$ &  116.5 & &  $7.8\pm2.7$ & -157.5 & &  $8.6\pm20.6$ & -176.7 \\
        & 15 &  $91,8\pm1322.3$ &   89.2 & &  $5.9\pm3.9$ & -150.8 & &  $5.6\pm17.0$ & -152.7 \\
        & 20 &  $119.1\pm1409.0$ &   83.9 & &  $12.8\pm143.5$ & 106.6 & &  $4.1\pm84.4$ & -173.7 \\
        & 25 & $141.1\pm1471.1$ &   69.7 & &  $57.0\pm364.9$ & 95.8 & &  $28.1\pm655.3$ &  100.7 \\
        & 30 &  $113.5\pm1565.4$ &   63.5 & &  $73.7\pm383.1$ &  94.9 & &  $57.3\pm1123.1$ &  96.1 \\ \hline
Per\,49 &  5 &  $11.7\pm48.6$ &   93.5 & &  $9.0\pm2.2$ &  140.5 & &  $7.3\pm14.5$ &   99.7 \\
        & 10 &   $9.0\pm43.2$ &  145.1 & &  $6.6\pm3.3$ &  144.3 & &  $8.5\pm14.1$ &  138.6 \\
        & 15 &   $4.2\pm250.4$ & 141.0 & &  $4.2\pm3.5$ &  138.9 & &  $5.7\pm15.3$ &  142.5 \\
        & 20 &  $55.3\pm1028.3$ &   -34.7 & &  $3.3\pm3.6$ &  131.6 & &  $2.8\pm166.2$ &  131.3 \\
        & 25 &  $102.5\pm1282.1$ &  -62.9 & &  $2.8\pm67.4$ &  58.3 & &  $4.7\pm582.0$ &  -78.4 \\
        & 30 &  $132.4\pm1401.4$ &  0.0 & &  $61.8\pm345.2$ &   -27.0 & &  $43.8\pm1204.0$ &   -10.4 \\ 
\enddata
\end{deluxetable*}

\begin{figure*}
    \centering
    \includegraphics[ width=0.98\textwidth, trim = 6mm 5mm 5mm 0mm, clip=True]{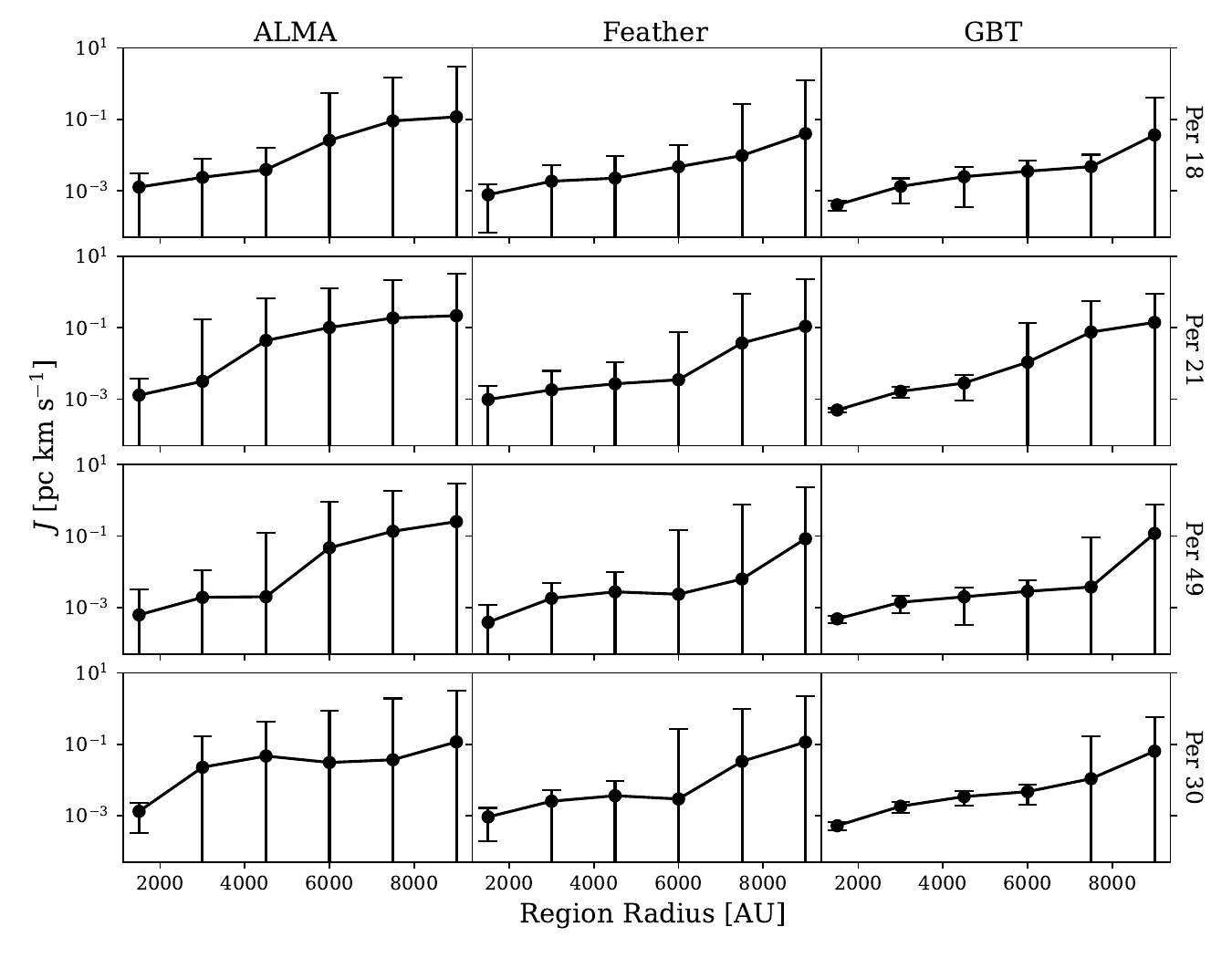}
    \caption{Specific angular momentum as a function of distance from the protostar for each of our four protostars and three image types.}
    \label{fig:specAngMom}
\end{figure*}

\begin{figure*}
    \centering
    \includegraphics[scale=0.74]{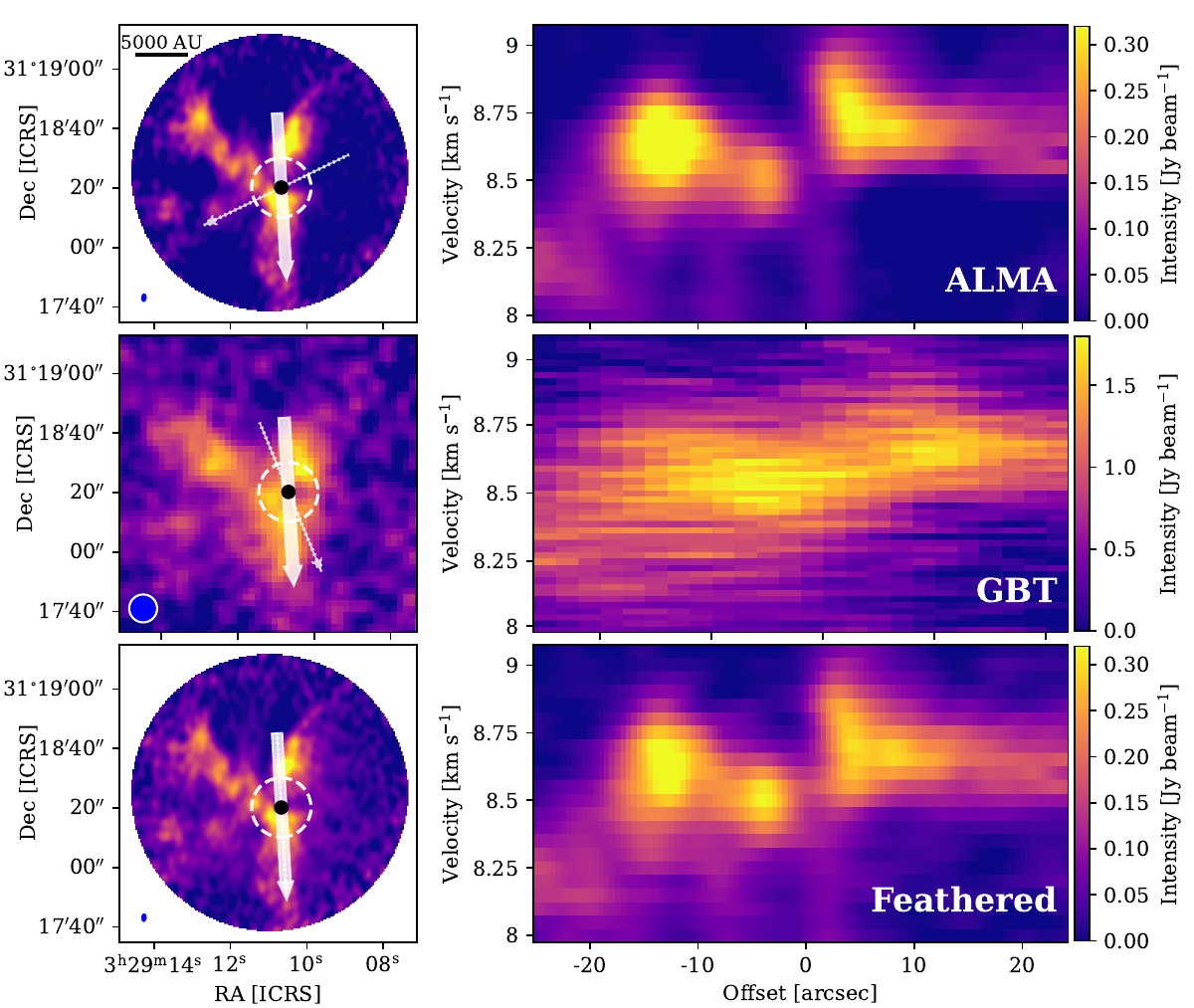} 
    \caption{PV diagrams for Per\,21 from ALMA (top), GBT (middle), and Feather (bottom) data. Left panels show data cube slices at $v=8.7$\,km\,s$^{-1}$. Thick white arrows correspond to the axes along which the PV diagrams are made, which are derived from the direction of $\nabla v_\text{lsr}$ calculated from the Feather data for a region with a 10$^{\prime\prime}$ radius (dashed white circle; see Table \ref{tab:lvgrad} for $\nabla v_\text{lsr}$). Thin white arrows are displayed as a reference and show the direction of the linear velocity gradient for the dataset in that row. Offset in the right column is positive in the direction of the arrows. Blue ellipses in bottom left corner of left-column panels represent the ALMA, GBT, and Feather beams.}
    \label{fig:per21_pvs}
\end{figure*}

\section{Discussion} \label{sec:discussion}

%\subsection{Linear Velocity Gradients, Angular Momentum, and Position-Velocity Diagrams} \label{sec:lvgrad+pv}

\subsection{Variation of Velocity Structure Across Spatial Scales} \label{sec:vel_variation}

We consider the implications of our calculated linear velocity gradients (Table \ref{tab:lvgrad}) and how they compare with similar studies. At the smaller scales ($\lesssim {10}^{\prime\prime}$), where our uncertainties are lower, $\nabla v_{\textrm{lsr}}$ is generally higher when examining the clumpier gas revealed in our ALMA data as compared to the more diffuse gas probed by the GBT. This finding may be a result of the velocity variation being averaged out in the lower-resolution GBT data, but it also could indicate that there is a greater change in gas velocity in clumpier gas versus diffuse gas within the regions of interest. In either case, these calculated $\nabla v_{\textrm{lsr}}$ values point to turbulent velocity structure in the smaller-scaled gas traced by ALMA.

We also find that $\nabla v_{\textrm{lsr}}$ trends downward with increasing radii in the GBT and Feather datasets, though this becomes less certain at radii $\gtrsim 15^{\prime\prime}$, particularly with the feather data. We do not see the same behavior in the ALMA data, which instead seem to feature little relationship between $\nabla v_{\textrm{lsr}}$ and radius. The ALMA $\nabla v_{\textrm{lsr}}$ measurements also feature a greater discrepancy between the angles of the linear velocity gradients at different radii. While $\theta_{\nabla v_{\textrm{lsr}}}$ for the GBT and Feather images vary by $\sim10-20$ degrees across radii for a given protostar (except in the case of Per\,49, which we discuss further below), our ALMA data feature linear velocity gradients that vary in direction by up to 90 degrees, indicating there is far less coherence in the velocity structure at different radii for small-scale versus large-scale material. Given that the Largest Angular Scale (LAS) for the ALMA measurements is $\sim33''$, equivalent in size to the 15$''$-radius aperture, we would not expect to probe large, coherent structure with ALMA-only measurements using apertures larger than $15''$ in radius. 

This constraint is likely a contributing factor to the very large uncertainties we calculate for the ALMA $\nabla v_{\text{lsr}}$ values listed in Table \ref{tab:lvgrad}, which represent the standard deviations of the velocity gradients within the given apertures. Though the uncertainties are larger than the ALMA $\nabla v_{\text{lsr}}$ values in nearly all cases, they are larger than the linear velocity gradients by up to two orders of magnitude for radii $>15''$. At these larger radii, we are likely averaging together several different velocity components, rather than  probing one component that spans the scale of the aperture, so the pixel-by-pixel velocity gradient scatter is large. 

Additionally, apertures with larger radii include more pixels from the edge of each image where emission is significantly lower in our ALMA measurements, often below the $5\,\sigma$ cutoff that was used when modeling the spectra. At these distances from the protostars, there is little dense, clumpy gas and thus the source emission is dominated by larger-scale gas to which ALMA is blind. As a result, blank pixels or those with poorer centroid velocity fits are included in the linear velocity gradient calculations, resulting in much greater scatter and larger uncertainties. Conversely, we see that the GBT $\nabla v_{\text{lsr}}$ uncertainties are on the same order of magnitude as the linear velocity gradients themselves, indicating we are probing much more coherent large-scale structure within a given aperture. The feather $\nabla v_{\textrm{lsr}}$ values are therefore in between the ALMA and GBT results, though they are more closely aligned with both the $\nabla v_{\textrm{lsr}}$ and position angle values calculated from the GBT data.

We note that the trends we identify in the velocity structure of the IRAS\,7 cores are also consistent with those seen in the isolated protostar Per\,30. Though one could argue that the contradictory ALMA linear velocity gradient values calculated for the IRAS\,7 protostars result from their complex velocity structure as part of a multiple-star system, we see the same behavior in Per\,30. Therefore, we posit that on the smaller spatial scales probed by ALMA, the velocity structure is not smooth and coherent around any of the studied protostars. The magnitudes of the linear velocity gradients calculated for Per\,30 from each dataset are also consistent with those found in the IRAS\,7 protostar. We do not see any significant differences in the linear velocity gradients between this isolated protostar and those from a multiple-protostar system.

We compare trends in our $\nabla v_{\textrm{lsr}}$ values to those noted in \cite{Sai2023}, which examined combined interferometric and single-dish measurements toward three protostars (IRAS\,15398-3359, L1527 IRS, and TMC-1A) in order to study the gas kinematics associated with the formation of protostars and their associated disks at sub-arcsecond resolution. \cite{Sai2023} used combined ALMA, IRAM-30 m, and Atacama Pathfinder Experiment (APEX) measurements of C$^{18}$O to fit the centroid velocity around these protostars and calculate velocity gradients at radii of 5--60$''$. Though C$^{18}$O may of course feature different kinematic behavior than \nth, \cite{Sai2023} found that the velocity gradients were larger nearer to the protostars and decreased with distance in all cases, which generally agrees with our findings for GBT and Feather data for radii $\lesssim 10-15^{\prime\prime}$. Though \cite{Sai2023} did not investigate the velocity gradients associated with just their ALMA data, we note that we find no discernible relationship between our ALMA-only velocity gradient values and distance from the protostar. These results suggests that the dense gas velocity structure is varying on arcsecond scales, while the velocity structure of the diffuse gas is varying on scales of tens of arcseconds or greater.

\vspace{1cm}

\subsection{Testing for Signs of Rotation} \label{sec:rotation}

Using our values for $\nabla v_{\textrm{lsr}}$ (Section \ref{sec:lvgrad}), we calculate the specific angular momentum $J_{\textrm{core}}$ for each region size and plot these for each protostar and datatype (ALMA-only, GBT-only, and Feather) in Figure \ref{fig:specAngMom}, which features profiles consistent with those seen in, for example, \cite{Goodman1993,Pineda2019} \citep[see][for a compilation of angular momentum calculations from other works]{Li2014}. It is clear that the specific angular momentum $J_{\textrm{core}}$ increases with distance from the protostar rather than remaining constant with radius. Thus, we can conclude that the velocity structure in \nth does not trace a circumstellar region of conserved angular momentum, as one may expect, for example, if the region is collapsing quickly with little time for angular momentum removal \citep[through, e.g., magnetic braking;][]{Li2014}. Such a region may still exist on a smaller scale closer to the protostar \citep[e.g.][]{Gaudel2020}. The trend of increasing $J_{\textrm{core}}$ with radius that we find in \nth is similar to that found by \cite{Pineda2019} \citep[see also][]{Caselli2002}, which they attribute, at least in part, to turbulence. This conclusion is consistent with those drawn from our centroid velocity maps, which also do not show clear signs of rotation in the IRAS\,7 protostar maps. 
%Additionally, Figure \ref{fig:specAngMom} demonstrates that the loss of angular momentum from larger to smaller radii is more pronounced in the denser gas traced by our ALMA data, suggesting that the motions of the dense gas are less influenced by rotation than those of the larger-scale, more diffuse gas \textcolor{red}{[need to confirm whether this conclusion is valid]}.

To further investigate the velocity structure of the \nth gas, we consider position-velocity (PV) diagrams made along the axes derived from our calculations of $\nabla v{_{\textrm{lsr}}}$ (see Table \ref{tab:lvgrad} for position angles). An example set of PV diagrams for Per\,21 using a $10''$-radius aperture is shown in Figure \ref{fig:per21_pvs}. Though the GBT PV diagram shows a smoother velocity structure, the smaller spatial scales probed in the ALMA and Feather diagrams reveal a much more fragmented picture, featuring several distinct clumps of gas traveling at different velocities. However, despite the differences between these PV diagrams across scales, we see no evidence for organized, coherent rotation on scales of $\sim1000$--2700\,AU, or 0.005--0.01\,pc. These results agree with findings from \cite{Sai2023}, which also determined that indications of rotation around protostars are not clear on scales $\gtrsim 1000$\,AU. Note that \cite{Tobin2018} measured the emission from several molecular tracers toward Per\,18 (see Section \ref{sec:intro}) with velocity gradients perpendicular to the outflow direction, indicating the kinematics they probe result from rotation. However, these measurements had a beam size of $0.74''\times0.31''$ (222\,AU\,$\times$\,93\,AU), again lending credence to the idea that we are not sensitive to rotation at the spatial scales we probe in this work. Higher resolution images of a greater number of protostars are necessary in order to determine at what scale rotation becomes apparent.

\subsection{Alternative Disk-Formation Mechanisms and Sources of Angular Momentum} \label{sec:alt_sources}

Though past work has suggested that the inheritance of rotation from a cloud or filament to protostellar scales is necessary in order for protostellar disks to form \citep{Shu1987}, our results indicate that angular momentum is not conserved across such a vast range of spatial scales \citep[consistent with, for example,][]{Pineda2019}. Thus we must consider alternative methods for disk formation as well as explanations for the velocity structure probed by our \nth measurements. \cite{Kuznetsova2019} consider numerical simulations from the \texttt{Athena} hydrodynamic code and find that the larger-scale rotation of a molecular cloud has little influence on the angular momentum of resulting cores. Instead, they suggest that torques as a result of nearby concentrations of mass are likely the cause of core-scale angular momentum. \cite{Verliat2020} follow a similar vein and propose non-axisymmetric collapse as an alternative disk formation mechanism that does not rely on initial cloud rotation. They present simulations which start with a dense core with an irregular density distribution and find that the asymmetry of the resulting collapse can form a rotating disk even without initial larger-scale rotation. The irregularities in the density distribution offset the collapse from the core's center of mass, producing torques that result in a rotating disk \citep{Pineda2023}. \cite{Verliat2020} also demonstrate that angular momentum will not be conserved in this formation scenario, which agrees with our findings.

\cite{Gaudel2020} also consider angular momentum across spatial scales. Upon examination of C$^{18}$O $J=2-1$ and \nth $J=1-0$ velocity gradients around 12 protostars from the IRAM CALYPSO survey \citep{Maury2014}, they find that their measurements support the presence of rotation in the inner envelope ($<1600$\,AU), as well as angular momentum conservation during core collapse. However, they find that the velocity gradients for the outer envelope ($>1600$\,AU) are not consistent with those nearer to the protostar, suggesting that different processes contribute to the velocity structure in the inner and outer envelope. Since the minimum radius we consider (5$''\sim1500$\,AU) when analyzing velocity structure is approximately equal to the radius at which \cite{Gaudel2020} find inconsistencies in their velocity gradients, it is possible that our resolution is too coarse to capture the rotation of the inner envelope, thus resulting in our measurements only reflecting the kinematic structure of the outer envelope without clear signatures of consistent rotation.

\section{Conclusion}

We examine the kinematic structure of two protostellar cores in the Perseus molecular cloud: the isolated protostar Per\,30 and the multiple-protostar system NGC\,1333 IRAS\,7. We use observations of \nth $J=1-0$ emission from ALMA and GBT and  combine these data in the Fourier plane to create images that are sensitive to all spatial scales larger than $\sim5^{\prime\prime}$. We use \nth\!\!\!'s hyperfine structure to examine the velocity gradients around the four protostellar systems within these two cores and find that the kinematic structures are not consistent with simple rotation. Analysis of the ALMA data reveals clumpy, turbulent gas structure on spatial scales of $\sim1500-3000$\,AU, whereas the more diffuse, larger-scale gas probed by the GBT on scales $\gtrsim 3000$\,AU demonstrates smoother velocity structure. 

These results are in agreement with those from the literature, which suggest that evidence of rotation is not apparent on scales larger than $\sim1000-1500$\,AU. Additionally, we suggest that the velocity structure and core angular momentum analyzed here have not been inherited from larger-scale progenitors (such as core- or cloud-scale components) in the star formation process. We find that both the isolated (Per\,30) and multiple-protostar (IRAS\,7) systems studied here support these results, indicating that the multiplicity of a system does not affect our ability to identify rotation in the velocity structure at the scales studied in this work. Instead we hypothesize that the chaotic velocity structure we measure results from asymmetrical or irregular density distributions, which can then cause torques, as well as contribute to protoplanetary disk rotation. Additional measurements and analysis of protostellar gas structure on scales $\lesssim1000$\,AU are needed in order to compare the disk- and core-scale velocity structures. Further work on the kinematics of multiple-protostar systems would also aid in illuminating the formation mechanisms of these complex, multi-star systems.

\section{Acknowledgments}
We thank C.Y. Chen for her work in kickstarting this project and obtaining the data. This paper makes use of the following ALMA data: ADS/JAO.ALMA\#2019.1.00914.S. ALMA is a partnership of ESO (representing its member states), NSF (USA) and NINS (Japan), together with NRC (Canada), MOST and ASIAA (Taiwan), and KASI (Republic of Korea), in cooperation with the Republic
of Chile. The Joint ALMA Observatory is operated by ESO, AUI/NRAO and NAOJ. The National Radio Astronomy Observatory and Green Bank Observatory are facilities of the U.S. National Science Foundation operated under cooperative agreement by Associated Universities, Inc. Support for this work was provided by the NSF through
the Grote Reber Fellowship Program administered by Associated Universities, Inc./National Radio
Astronomy Observatory. ZYL is supported in part by NASA 80NSSC20K0533, NSF AST-2307199, and the Virginia Institute of Theoretical Astronomy (VITA).

\newpage
\appendix 

\section{Spectral Fits} \label{sec:allFits}

We provide here spectral fits from \texttt{PySpecKit} toward the remaining three protostars (Figures~\ref{fig:specFit_Per30}, \ref{fig:specFit_Per21}, and \ref{fig:specFit_Per49}) using the methods described in Section~\ref{sec:specFit}.

\begin{figure}
    \gridline{\fig{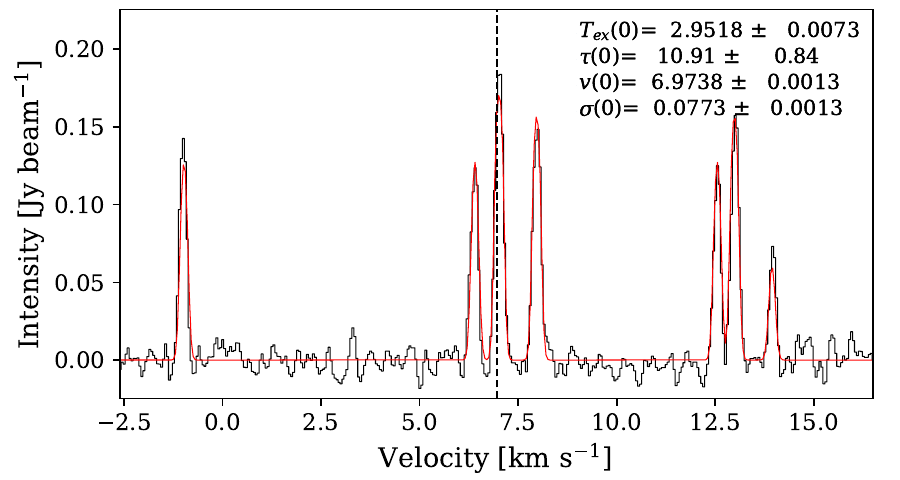}{0.6\textwidth}{(a)}}
    \vspace{-3mm}
    \gridline{\fig{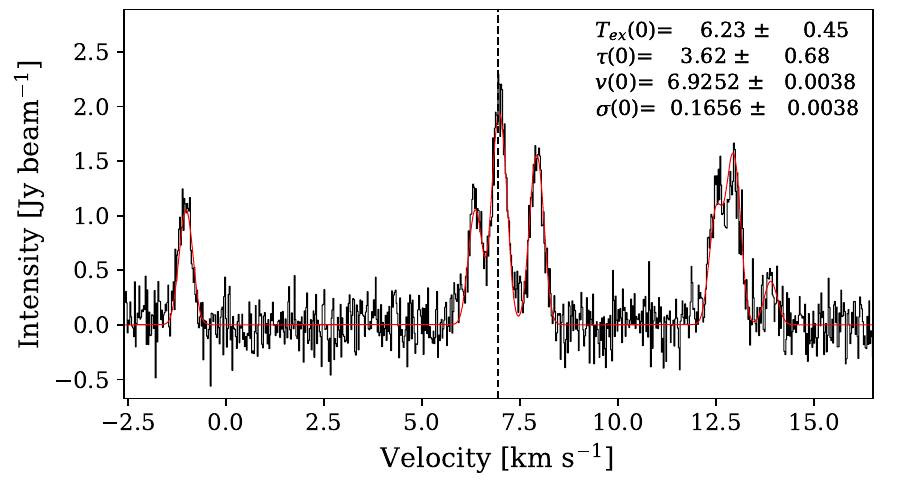}{0.6\textwidth}{(b)}}
    \vspace{-3mm}
    \gridline{\fig{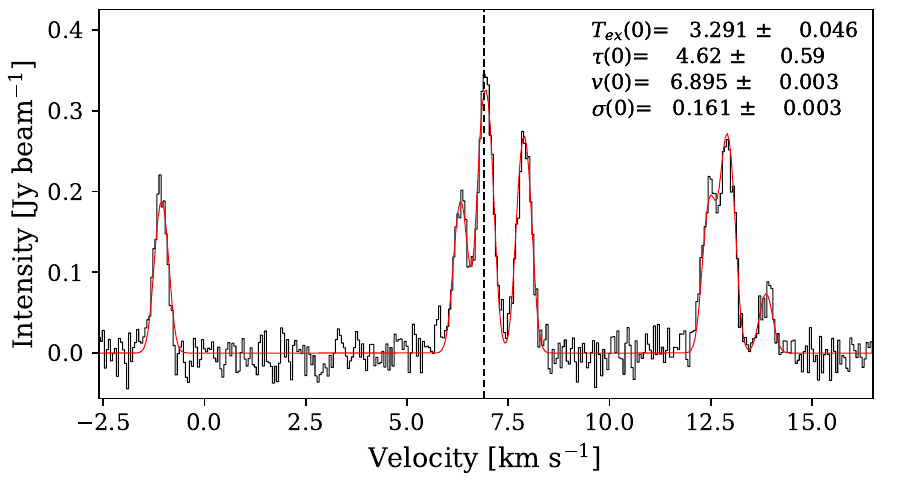}{0.6\textwidth}{(c)}}
    \vspace{-3mm}
    \caption{Sample \nth $J=1-0$ spectral fits (red curves) of the ALMA (a), GBT (b), and Feather (c) spectra (solid black curves) toward the ALMA Per\,30 continuum peak. Each panel shows the best-fit values and uncertainties for excitation temperature $T_\text{ex}$ in K, the total optical depth $\tau$ (which is distributed across the 15 hyperfine components), centroid velocity $v$ (also denoted by the dashed black lines) in km\,s$^{-1}$, and linewidth $\sigma$ in km\,s$^{-1}$.}
    \label{fig:specFit_Per30}
\end{figure}

\begin{figure}
    \gridline{\fig{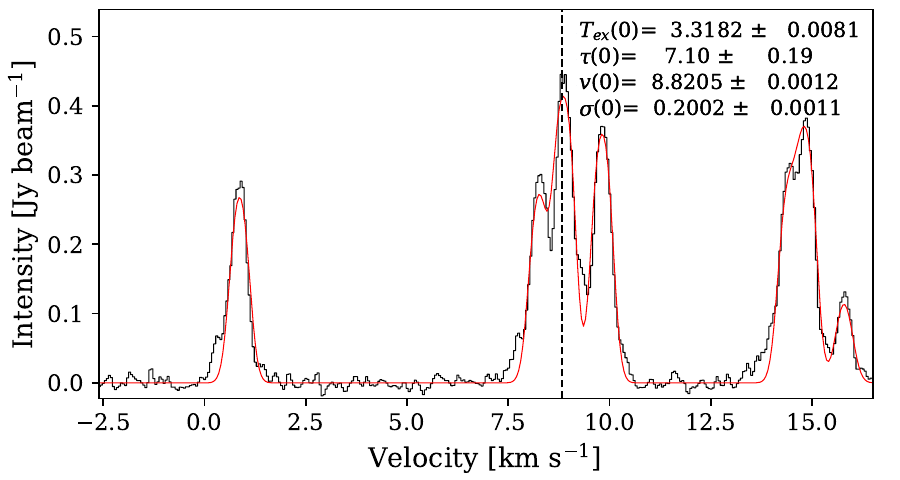}{0.6\textwidth}{(a)}}
    \vspace{-3mm}
    \gridline{\fig{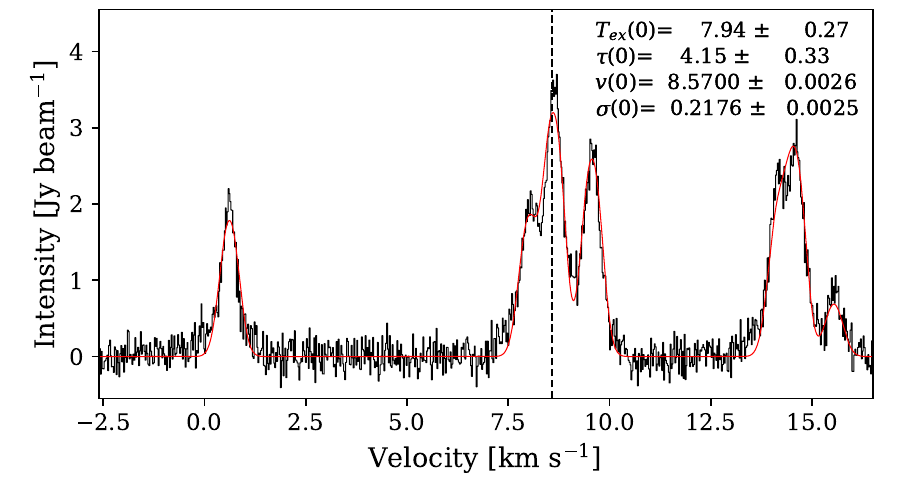}{0.6\textwidth}{(b)}}
    \vspace{-3mm}
    \gridline{\fig{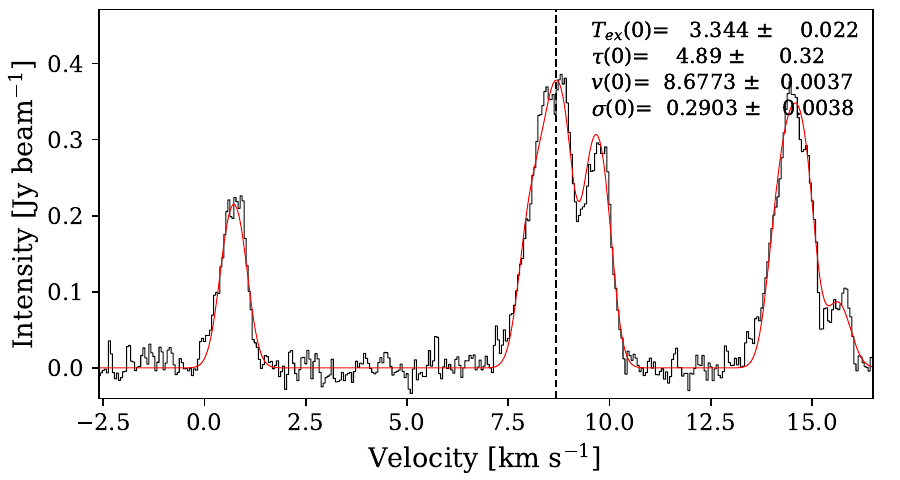}{0.6\textwidth}{(c)}}
    \vspace{-3mm}
    \caption{Sample \nth $J=1-0$ spectral fits (red curves) of the ALMA (a), GBT (b), and Feather (c) spectra (solid black curves) toward the ALMA Per\,21 continuum peak. Each panel shows the best-fit values and uncertainties for excitation temperature $T_\text{ex}$ in K, the total optical depth $\tau$ (which is distributed across the 15 hyperfine components), centroid velocity $v$ (also denoted by the dashed black lines) in km\,s$^{-1}$, and linewidth $\sigma$ in km\,s$^{-1}$.}
    \label{fig:specFit_Per21}
\end{figure}

\begin{figure}
    \gridline{\fig{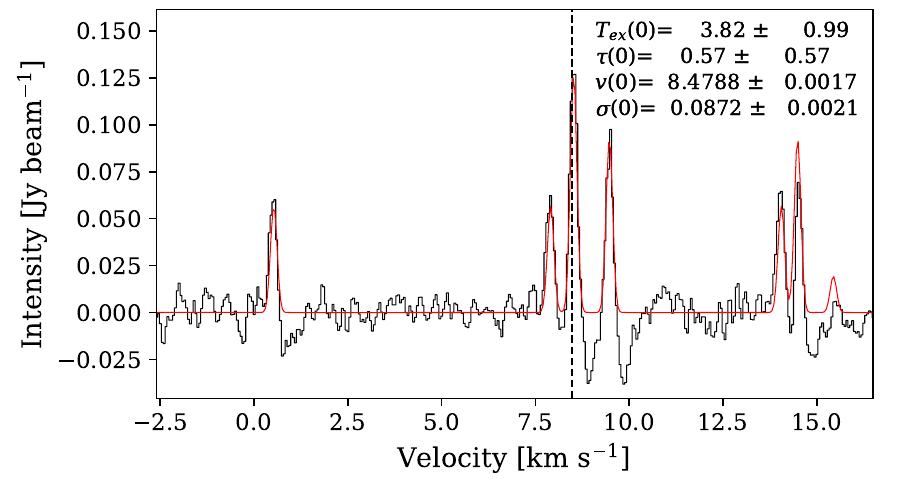}{0.6\textwidth}{(a)}}
    \vspace{-3mm}
    \gridline{\fig{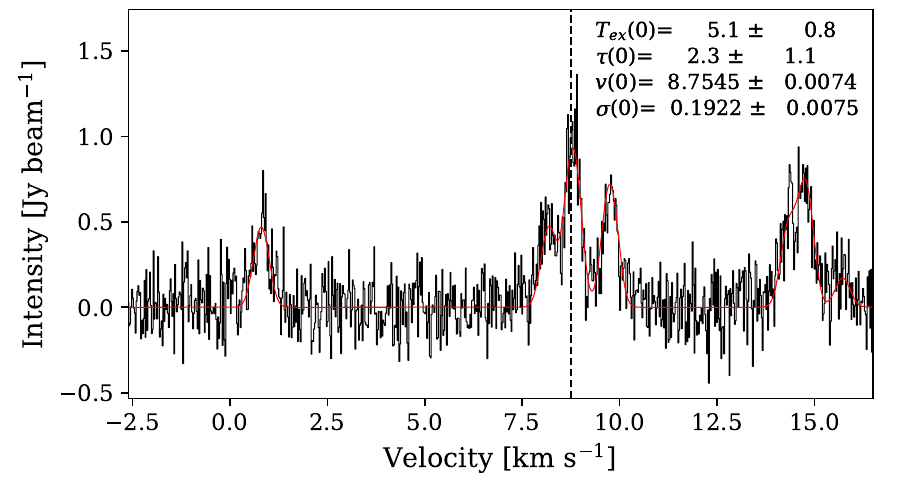}{0.6\textwidth}{(b)}}
    \vspace{-3mm}
    \gridline{\fig{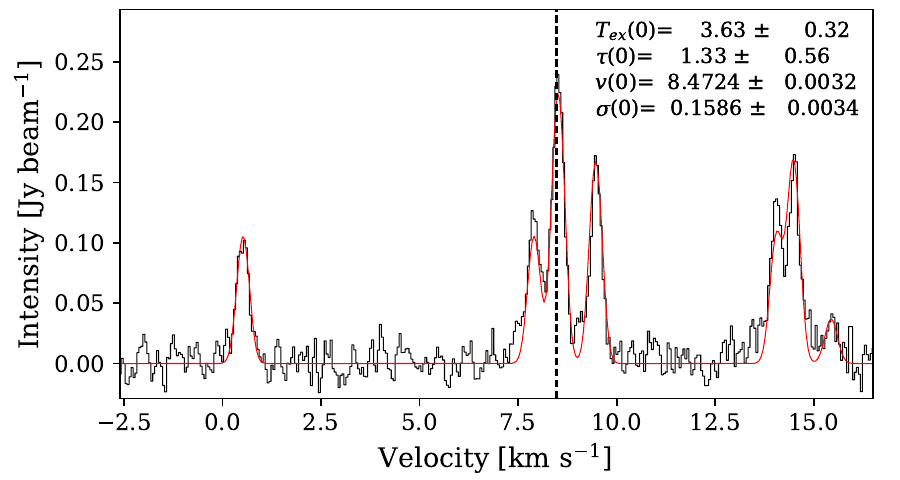}{0.6\textwidth}{(c)}}
    \vspace{-3mm}
    \caption{Sample \nth $J=1-0$ spectral fits (red curves) of the ALMA (a), GBT (b), and Feather (c) spectra (solid black curves) toward the ALMA Per\,49 continuum peak. Each panel shows the best-fit values and uncertainties for excitation temperature $T_\text{ex}$ in K, the total optical depth $\tau$ (which is distributed across the 15 hyperfine components), centroid velocity $v$ (also denoted by the dashed black lines) in km\,s$^{-1}$, and linewidth $\sigma$ in km\,s$^{-1}$.}
    \label{fig:specFit_Per49}
\end{figure}


\newpage
\bibliography{N2H+}{}
\bibliographystyle{aasjournal}

%\restartappendixnumbering
%\appendix 

%
%\end{CJK*}
\end{document}